\documentclass[]{emulateapj}
\usepackage{epsfig, natbib, graphicx, color}

\input epsf 

\usepackage{color}

\newcommand{\msun}{M_\odot}
\newcommand{\hmsun}{h^{-1}\ M_\odot}
\newcommand{\bm}[1]{\mathbf{#1}}

\newcommand{\nn}{\nonumber}

\newcommand{\nM}{ \frac{dn}{dM}}

\newcommand{\avg}[1]{\left\langle #1 \right\rangle}
\newcommand{\Var}{\mbox{Var}}

\newcommand{\lk}{{\cal{L}}}

\newcommand{\LCDM}{\Lambda\mbox{CDM}}

\newcommand{\lkhd}{\mbox{$\cal{L}$}}

\newcommand{\bx}{\bm{x}}

\newcommand{\Lx}{L_X}

\shortauthors{Rozo et al.}
\shorttitle{Cosmological Constraints from maxBCG Clusters}

\begin{document}
\title{Cosmological Constraints from the SDSS maxBCG Cluster Catalog}
\author{Eduardo Rozo\altaffilmark{1}, 
Risa H. Wechsler\altaffilmark{2},
Eli S. Rykoff\altaffilmark{3}, 
James T. Annis\altaffilmark{4}, 
Matthew R. Becker\altaffilmark{5,6},
August E. Evrard\altaffilmark{7,8,9}, 
Joshua A. Frieman\altaffilmark{4,6,10}
Sarah M. Hansen\altaffilmark{11}, 
Jiangang Hao\altaffilmark{7}, 
David E. Johnston\altaffilmark{12}, 
Benjamin P. Koester\altaffilmark{6,10},
Timothy A. McKay\altaffilmark{7,8,9}, 
Erin S. Sheldon\altaffilmark{13},
David H. Weinberg\altaffilmark{1,14}
}
\altaffiltext{1}{Center for Cosmology and Astro-Particle Physics, The Ohio State University, Columbus, OH 43210 {\tt erozo@mps.ohio-state.edu}}
\altaffiltext{2}{Kavli Institute for Particle Astrophysics \& Cosmology,
  Department of Physics, and SLAC National Accelerator Laboratory
  Stanford University, Stanford, CA 94305}
\altaffiltext{3}{TABASGO Fellow, Physics Department, University of California at Santa Barbara, 2233B Broida Hall, Santa Barbara, CA 93106}
\altaffiltext{4}{Fermi National Accelerator Laboratory, P.O. Box500, Batavia, IL 60510}
\altaffiltext{5}{Department of Physics, The University of Chicago, Chicago, IL 60637}
\altaffiltext{6}{Kavli Institute for Cosmological Physics, The University of Chicago, Chicago, IL 60637} 
\altaffiltext{7}{Physics Department, University of Michigan, Ann Arbor, MI 48109}
\altaffiltext{8}{Astronomy Department, University of Michigan, Ann Arbor, MI 48109}
\altaffiltext{9}{Michigan Center for Theoretical Physics, Ann Arbor, MI 48109}
\altaffiltext{10}{Department of Astronomy and Astrophysics, The University of Chicago, Chicago, IL 60637}
\altaffiltext{11}{University of California Observatories \& Department of Astronomy, University of California, Santa Cruz, CA 95064}
\altaffiltext{12}{Department of Physics \& Astronomy, Northwestern University, 
Evanston, IL 60208}
\altaffiltext{13}{Brookhaven National Laboratory, Upton, NY 11973}
\altaffiltext{14}{Department of Astronomy, The Ohio State University, Columbus, OH 43210}

\begin{abstract}
  We use the abundance and weak lensing mass measurements of the SDSS
  maxBCG cluster catalog to simultaneously constrain cosmology and the
  richness--mass relation of the clusters.  Assuming a flat
  $\Lambda$CDM cosmology, we find $\sigma_8(\Omega_m/0.25)^{0.41} =
  0.832\pm 0.033$ after marginalization over all systematics.  
  In common with previous studies, our error budget is dominated by systematic
  uncertainties, the primary two being the absolute mass scale of the weak lensing
  masses of the maxBCG clusters, and uncertainty in the
  scatter of the richness--mass relation.
  Our constraints are
  fully consistent with the WMAP five-year data, and in
  a joint analysis we find $\sigma_8=0.807\pm 0.020$ and
  $\Omega_m=0.265\pm 0.016$, an improvement of nearly a factor of two
  relative to WMAP5 alone.  
  Our results are also in excellent agreement
  with and comparable in precision to the latest cosmological
  constraints from X-ray cluster abundances.  The remarkable
  consistency among these results demonstrates that cluster
  abundance constraints are not only tight but also robust, and
  highlight the power of optically-selected cluster samples to
  produce precision constraints on cosmological parameters.
\end{abstract}

\section{Introduction}

The abundance of galaxy clusters has long been recognized as a
powerful tool for constraining cosmological parameters.  More
specifically, from theoretical considerations
\citep[e.g.][]{pressschechter74,bondetal91,whiteetal93,shethtormen02} one expects
the abundance of massive halos to be exponentially sensitive to the
amplitude of matter fluctuations.     Though some theoretical challenges
remain \citep[see e.g.][]{robertsonetal08,staneketal09}, 
this basic theoretical prediction 
has been confirmed many times in detailed numerical simulations, 
and a careful calibration of the abundance of halos as a function of mass
for various cosmologies has been performed \citep[see
e.g.][]{jenkinsetal01, warrenetal06, tinkeretal08}. 
Despite these
successes, realizing the promise of cluster cosmology has proven
difficult.  Indeed, a review of observational results from the past
several years yields a plethora of studies where typical
uncertainties are estimated at the $\Delta\sigma_8\approx 0.05-0.10$
level despite a spread in central values that range from
$\sigma_8\approx 0.65$ to $\sigma_8 \approx 1.0$
\citep[][]{vianaliddle96,vianaliddle99,
  henry91,henry00,pierpaolietal01,borganietal01,
  seljak02,vianaetal02,
  schueckeretal03,allenetal03,bahcalletal03,bahcallbode03,
  henry04,
  voevodkinvikhlinin04,rozoetal07a,gladdersetal07,rinesetal07}.

The discrepancies among the various studies mentioned above is a
manifestation of the fundamental problem confronting cluster abundance
studies: theoretical predictions tell us how to compute the abundance
of halos as a function of mass, but halo masses are not observable.
Consequently, we are forced to rely on observable quantities such as
X-ray temperature, weak lensing shear, or other such signals, to estimate cluster
masses.  This reliance on observable mass tracers introduces
significant systematic uncertainties in the analysis; indeed, this is typically the
dominant source of error \citep[e.g.][]{henryetal08}.

There are two primary ways in which these difficulties can be
addressed.  One possibility is to reduce these systematic
uncertainties through detailed follow-up observations of relatively
few clusters, an approach exemplified in the work of
\citet{vikhlininetal08}.  The second possibility is to use large cluster samples
complemented with statistical properties of the clusters that are
sensitive to mass to simultaneously fit for cosmology and the observable--mass
relation of the cluster sample in question.   Indeed, this is the basic 
idea behind the so called self-calibration approach, in which one uses
the clustering of clusters \citep[][]{schueckeretal03,estradaetal08}
and cluster abundance
data to derive cosmological constraints with no a-priori knowledge
of the observable--mass relation  \citep[][]{hu03,majumdaretal04,limahu04,limahu05}.
There are, however, many other statistical observables that correlate well
with mass, such as the cluster--shear
correlation function \citep{sheldonetal07}, or even counts binned in
multiple mass tracers \citep{cunha08}.  By including such data we can
break the degeneracy between cosmology and the observable--mass
relation, thereby obtaining tight cosmological constraints while
simultaneously fitting the observable--mass relation.  

In this work, we derive
cosmological constraints from the SDSS maxBCG cluster sample
\citep{koesteretal07a} and the statistical weak lensing mass
measurement from \citet{johnstonetal07}.  We then compare our result
to three state-of-the-art cluster abundance studies of X-ray selected
cluster samples \citep{mantzetal08,henryetal08, vikhlininetal08b} and
demonstrate that our results are both consistent and
competitive with these studies.  This is the first time an optically
selected catalog with masses estimated in a statistical way
has produced constraints that are of
comparable accuracy to the more traditional approach.

The paper is organized as follows.  Section \ref{sec:data} presents
the data used in our study.  Section \ref{sec:analysis} describes our
analysis, including the likelihood model and priors adopted in this
work, and the way in which the analysis was implemented.  Section
\ref{sec:results} presents our main results, while sections
\ref{sec:observational_systematics}, \ref{sec:prior_systematics}, and
\ref{sec:parsys} discuss various sources of systematic uncertainties.
Section \ref{sec:comparison} compares our results to the most recent
results from X-ray selected cluster samples. Section \ref{sec:w}
investigates the implications of our results for dark energy, and
Section \ref{sec:improvement} discusses the prospects for improving
our cosmological constraints from the maxBCG cluster sample in the
future.  Section \ref{sec:summary} summarizes our main results and
conclusions.  Unless otherwise stated, all masses in this work are
defined using an overdensity $\Delta=200$ relative to the mean matter
density of the universe.


\section{Data}
\label{sec:data}

\subsection{MaxBCG Cluster Counts}

The maxBCG cluster catalog \citep{koesteretal07a} is an optically
selected catalog drawn from 7,398 $\deg^2$ of DR4+ imaging data of the Sloan Digital Sky
Survey (SDSS).\footnote{We write DR4+ as the catalog used a few hundred degrees
of imaging beyond those released with DR4.}   
The maxBCG algorithm exploits the tight E/S0 ridgeline
of galaxies in color-magnitude space to identify spatial overdensities of bright
red galaxies. 
The tightness of the color
distribution of cluster galaxies greatly suppresses the projection
effects that have plagued optically selected cluster
catalogs, and also allows for accurate photometric redshift estimates
of the clusters ($\Delta z \approx 0.01$).  MaxBCG clusters are
selected such that their photometric redshift estimates are in the range
$z_{photo}\in[0.1,0.3]$, resulting in a nearly volume-limited catalog.
A detailed discussion of the maxBCG cluster finding algorithm can be
found in \citet{koesteretal07}.

We bin the maxBCG cluster sample in nine richness bins spanning the range
$N_{200} \in[11,120]$, corresponding roughly to $M\in[7\times10^{13}
\hmsun, 1.2 \times 10^{15} \hmsun]$.  Our richness measure $N_{200}$ 
is defined as the number
of red-sequence galaxies within a scaled radius 
such that the average galaxy
overdensity interior to that radius is $200$ times the mean galaxy density of the universe
\citep[see][for further details]{koesteretal07a}.
The richness bins, and the
number of clusters in each bin, are presented in Table
\ref{tab:abundance_bins}.  There are an additional five clusters with
richness $N_{200}>120$.  These five clusters have $N_{200}=126$, $139$,
$156$, $164$, and $188$, and are properly included in the analysis on
an individual basis (see \S \ref{sec:lkhd} for details).


\begin{deluxetable}{c c}
\tablewidth{0pt}
\tablecaption{\label{tab:abundance_bins} Abundance of maxBCG Clusters}
\tablehead{Richness & No. of Clusters}
\startdata
11-14 & 5167 \\
14-18 & 2387 \\
19-23 & 1504 \\
24-29 & 765 \\
30-38 & 533 \\
39-48 & 230 \\
49-61 & 134 \\
62-78 & 59 \\
79-120 & 31\\
\enddata
\end{deluxetable}

\begin{figure}[t]
\epsscale{1.2}
\plotone{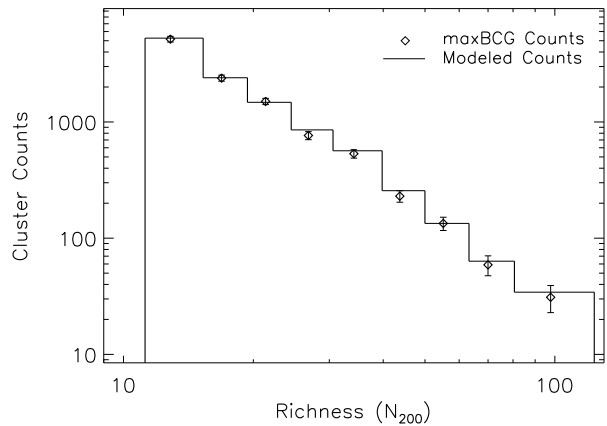}
\caption{Observed (diamonds) and modeled (solid line) cluster counts as a function 
  of richness
  in our best-fit model described in Section \ref{sec:results}.
  The model counts are computed using the best fit model detailed in
  Section \ref{sec:results}, and are a good fit to the data.}
\label{fig:obs_counts}
\end{figure}

Figure~\ref{fig:obs_counts} shows the cluster counts corresponding to
Table~\ref{tab:abundance_bins}.  Error bars between the various points
are correlated.   Also shown are the modeled counts from our best-fit
model, detailed in Section \ref{sec:results}.  We show these model
counts here for comparison purposes.


\subsection{MaxBCG Weak Lensing Masses}
\label{sec:masses}

\begin{figure}[t!]
\epsscale{1.2}
\plotone{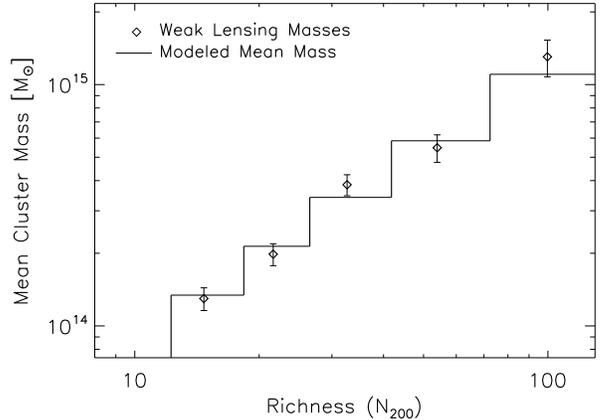}
\caption{Mean weak lensing mass of maxBCG clusters as a function
of richness.  The diamonds with error bars correspond to our data,
while the solid line shows the values predicted from our best-fit model
(see Section \ref{sec:results} for details).
We note the error bars are correlated, and the model is a good fit to the data.}
\label{fig:obs_masses}
\end{figure}

Estimates of the mean mass of the maxBCG clusters as a function of
richness are obtained through the weak lensing analysis described by
\citet{sheldonetal07} and \citet{johnstonetal07}.  Briefly,
\citet{sheldonetal07} binned the maxBCG cluster sample in richness
bins as summarized in Table~\ref{tab:mass_bins}.  Given a cluster in a
specified richness bin, they use all cluster--galaxy pairs with the
selected cluster as a lens to estimate the density contrast profile
$\Delta\Sigma$ of the cluster.  While these individual cluster
profiles are very low signal-to-noise, averaging over all clusters
within a richness bin allows one to obtain accurate estimates for the
mean density contrast profile of maxBCG clusters as a function of
richness.  The resulting profiles are fit using a halo model formalism
to derive mean cluster masses by \cite{johnstonetal07}.  We then
correct these masses upward by a factor of $1.18$ due to the expected
photometric redshift bias due to the dilution of the lensing signal
from galaxies that are in front of the cluster lenses, but whose
photometric redshift probability distribution extends past the cluster
lens \citep[see][for details]{mandelbaumetal08}.  A very similar but
independent analysis has also been carried out by
\citet{mandelbaumetal08b}, and we use the comparison between the two
independent analysis to set the systematic error uncertainty of the
weak lensing mass estimates \citep{rozoetal08a}.  The final results of
the weak lensing analysis summarized above are presented here in Table
\ref{tab:mass_bins}.\footnote{The number of clusters in Table
  \ref{tab:mass_bins} is larger than that reported in
  \citet{johnstonetal07} due to masking in the weak lensing
  measurements.  This additional masking does not bias the recovered
  masses in any way.}  Figure~\ref{fig:obs_masses} shows the mean weak
lensing masses from Table~\ref{tab:mass_bins}.  Also shown are the
mean masses computed using the best-fit model detailed in Section
\ref{sec:results}.  The richness binning of the weak lensing mass
estimates differs from that of the abundance data because of the
larger number of clusters necessary within each richness bin to obtain
high S/N weak lensing measurements.


\begin{deluxetable}{ccc}
\tablewidth{0pt}
\tablecaption{\label{tab:mass_bins} Mean Mass of maxBCG Clusters}
\tablehead{Richness & No. of Clusters & $\avg{M_{200b}} [ 10^{14}\
  \msun ]$}
\startdata
12-17 & 5651 & 1.298\\
18-25 & 2269 & 1.983\\
26-40 & 1021 & 3.846\\
41-70 & 353 & 5.475 \\
71+ & 55 & 13.03 
\enddata
\tablecomments{ Masses listed here are based on those quoted in
  \citet{johnstonetal07}, rescaled by the expected photometric
  redshift bias described in the text, and extrapolated to a matter
  overdensity $\Delta=200$ from the $\Delta=180$ value quoted in
  \citet{johnstonetal07}.  The masses have also been rescaled to the
  cosmology that maximizes our likelihood function, $(\sigma_8=0.80$,
  $\Omega_m=0.28$).}
\end{deluxetable}



\section{Analysis}
\label{sec:analysis}

We employ a Bayesian approach for deriving cosmological constraints
from the maxBCG cluster sample.  We use only minimal priors
placed on the parameters governing the richness--mass relation, relying
instead on the cluster abundance and weak lensing data to simultaneously
constrain cosmology and the richness--mass relation of the clusters.
Details of the model, parameter priors, and implementation can be
found below.

\subsection{Likelihood Model}
\label{sec:lkhd}

The observable vector $\bx$ for our experiment is comprised of:
\begin{enumerate}
\item $N_1$ through $N_{9}$: the number of clusters in each of the nine
  richness bins defined in Table \ref{tab:abundance_bins}.
\item $(N\bar M)_1$ through $(N\bar M)_5$, the total mass contained in
  clusters in each of the five richness bins defined in Table
  \ref{tab:mass_bins}, computed assuming $\Omega_m=0.27$ and 
  $h=0.71$.\footnote{While Table \ref{tab:mass_bins} reports the masses
  after corrections,  assuming $\Omega_m=0.28$, the actual input
  to our statistical analyis are the uncorrected masses from \citet{johnstonetal07},
  which assume $\Omega_m=0.27$.}
\end{enumerate}

We adopt a Gaussian likelihood model, which is fully specified by the
mean and covariance matrix of our observables.  Expressions for these
quantities as a function of model parameters are specified below. We
also multiply this Gaussian likelihood by a term that allows us to
properly include the information contained in clusters with richness
$N_{200}>120$.  In this richness range clusters are very rare and a
Gaussian likelihood model is not justified.  Instead we adopt a
likelihood model where the probability of having a cluster of a
particular richness $N_{200}$ is binary (i.e. a Bernoulli distribution), with
\begin{equation}
P(N|N_{200}) = \left\{ \begin{array}{cl}
	1-p & \mbox{if $N=0$} \\
	p & \mbox{if $N=1$}.
	\end{array} \right .
\end{equation}

Such a probability distribution is adequate so long as the probability
of having two clusters of a given richness is infinitesimally small.
Note that given this binary probability distribution, we have that the
expectation value of the number of such clusters is simply
$\avg{N(N_{200})}=p$, and the likelihood is fully specified by the
expectation value of our observable.  We find that the likelihood of
observing the particular richness distribution found for the maxBCG
catalog for clusters of richness $N_{200}\geq 120$ is
\begin{equation}
\lkhd_{tail} = \prod_{N(N_{200})=0}(1-\avg{N(N_{200})})\prod_{N(N_{200})=1} \avg{N(N_{200})}.
\end{equation}
%


The first product is over all richness $N_{200}>120$ and no clusters in
them, and the second product is over richness bins which contain one
cluster.  The subscript $tail$ reflects the fact that it is the
likelihood of the tail of the abundance function.  The final
likelihood $\lkhd=\lkhd_G\lkhd_{tail}$ is the product of the Gaussian
likelihood $\lkhd_G$ described earlier and the likelihood of the
abundance function tail.  We note that the log-likelihood of the tail
simplifies to
\begin{eqnarray}
\ln \lkhd_{tail} & = & \sum_{N_{200}>120} \avg{N(N_{200})}  \nn \\
 & & \hspace{0.15 in} - \sum_{N(N_{200})=1} \avg{N(N_{200})}+\ln \avg{N(N_{200})}.
\end{eqnarray}
An identical result is obtained assuming only Poisson variations in
the number of clusters for $N_{200}>120$.


\subsection{Expectation Values}
\label{sec:expectation_values}

To fully specify our likelihood model we need to derive expressions
for the mean and variance of our observables.  The model adopted in
this work is very similar in spirit to that of \citet{rozoetal07a}, so
we present here only a brief overview of the formalism.  Interested
readers can find a detailed discussion in \citet{rozoetal07a}.

We begin by considering the expected mean number of clusters in our
sample.  The number of halos within a redshift bin
$z\in[z_{min},z_{max}]$ and within a mass range $[M_{min},M_{max}]$ is
given by
\begin{equation}
N = \int dM\ dz\ \nM \frac{dV}{dz} \psi(M) \phi(z),
\label{eq:mass_cts}
\end{equation}
where $dn/dM$ is the halo mass function, $dV/dz$ is the comoving
volume per unit redshift, and $\psi(M)$ and $\phi(z)$ are the mass and
redshift binning functions. i.e. $\psi(M)=1$ if $M$ is within the mass
bin of interest and zero otherwise, and $\phi(z)=1$ if $z$ is within
the redshift bin of interest, but is zero otherwise.

\begin{figure}[t]
\epsscale{1.2}
\plotone{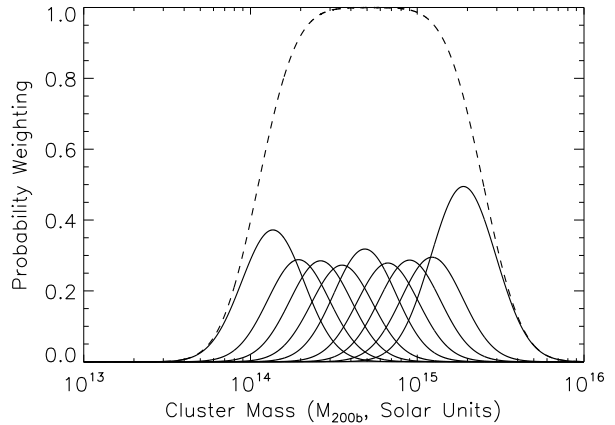}
\caption{ Mass selection function of the maxBCG algorithm.  The nine
  solid curves represent the probability that a halo of the
  corresponding mass falls within each of the nine richness bins
  described in Table \ref{tab:abundance_bins}. The dashed line 
  is the sum of all the binning
  functions, and is the probability that a halo of a given mass is
  assigned a richness $N_{200}\in [11,120]$, i.e.
  it is the mass selection function of the maxBCG algorithm over this
  richness range.  These binning functions are all estimated using our 
  best-fit model parameters, which are detailed in Section \ref{sec:results}.}
\label{fig:binning}
\end{figure}

In practice, we observe neither a cluster's mass nor its true
redshift, but are forced to rely on the cluster richness $N_{200}$ as a
mass tracer and to employ a photometric redshift estimate.  Let then
$P(N_{200}|M)$ denote the probability that a cluster of mass $M$ has a
richness $N_{200}$, and let $P(z_{photo}|z)$ denote the probability that
a cluster at redshift $z$ is assigned a photometric redshift
$z_{photo}$.  The binning function $\psi$ is now a function of
richness rather than mass so $\psi(N_{200})=1$ for
$N_{200}\in[N_{200}^{min},N_{200}^{max}]$.  Likewise, the redshift binning
function is now a function of photometric redshift $z_{photo}$.  The
total number of clusters in the maxBCG catalog becomes
\begin{equation}
\avg{N} = \int dM\ dz\ \nM \frac{dV}{dz} \avg{\psi|M} \avg{\phi|z},
\label{eq:mean_cts}
\end{equation}
where
\begin{eqnarray}
\avg{\psi|M} & = & \int dN_{200}\ P(N_{200}|M) \psi(N_{200}) \\
\avg{\phi|z} & = & \int dz_{photo}\ P(z_{photo}|z) \phi(z_{photo}). \label{eq:zbin}
\end{eqnarray}
The quantity $\avg{\psi|M}$ represents the probability that a halo of
mass $M$ falls within the richness bin defined by $\psi$.  We show
these probabilities as a function of mass for each of the nine
richness bins considered here in Figure~\ref{fig:binning}.  To make
the figure, we have set all relevant model parameters to their best-fit
value detailed in Section \ref{sec:results}.


A similar argument allows us to write an expression for the
expectation value for the total mass contained in clusters of a
specified richness and redshift bin.  This is given by
\begin{equation}
\avg{N\bar M} = \int dM dz \nM \frac{dV}{dz} M \avg{\psi|M}\avg{\phi|z}.
\label{eq:mass_prediction}
\end{equation}
The notation $N\bar M$ reflects the fact that if $\bar M$ is the mean
mass of the clusters of interest, the total mass contained in such
clusters is $N\bar M$ where $N$ is the total number of clusters in
said bin.

So far, our formulae adequately describe our experiment provided the
weak lensing masses estimated by \citet{johnstonetal07} are fair
estimates of the mean mass of the maxBCG clusters.  In practice, there
is an important systematic that needs to be properly incorporated in
our analysis, and which slightly modifies our expression.
We are referring to uncertainties
in the photometric redshift estimates of the source galaxies employed
in the weak lensing analysis.  The main problem here is that the mean
surface mass density profile $\Sigma(R)$ recovered by the weak lensing
analysis is proportional to $1/\avg{\Sigma_c^{-1}}$, the average
inverse critical surface density of all lens--source pairs employed in
the analysis.  We introduce an additional weak lensing bias parameter
$\beta$ such that if $\bar M_{true}$ is the true mean mass of a set of
clusters, the weak lensing mass estimate $\bar M_{obs}$ is given by
$\bar M_{obs} = \beta \bar M_{true}$.  Consequently, our final
expression for the mean weak lensing masses of the maxBCG clusters is
\begin{equation}
\avg{N\bar M} = \beta \int dM dz\ \nM \frac{dV}{dz} M \avg{\psi|M}\avg{\phi|z}.
\label{eq:mass_prediction_final}
\end{equation}
Priors on the parameter $\beta$ are discussed in Section \ref{sec:priors}.


\subsection{Covariance Matrix}

There are multiple sources of statistical uncertainty in the data.
These include: (1) Poisson fluctuations in the number of halos of a
given mass, (2) variance in the mean overdensity of the survey volume,
and (3) fluctuations in the number of clusters at fixed richness due to
stochasticity of the richness--mass relation.  The covariance matrix
of the observables is defined by the sum of the covariance matrices
induced by each of the three sources of statistical fluctuation just
mentioned.  A detailed derivation of the relevant formulae is
presented in \citet{rozoetal07b}.  Since this derivation generalizes
trivially to include the mean mass as an additional observable --- one
needs only to introduce a mass weight in the formulae as appropriate
--- we will not repeat ourselves here.

There is, however, one additional source of statistical uncertainty
that is not included in these calculation, namely measurement error in
the weak lensing masses.  More specifically, uncertainties in the
recovered weak lensing masses is dominated by shape noise in the
source galaxies.  This error was estimated by \citet{sheldonetal07}
using jackknife resampling, and was properly propagated into the
computation of the weak lensing mass estimates by
\citet{johnstonetal07}.  This error is added in quadrature to the
diagonal elements of the covariance matrix corresponding to the mean
mass measurements.

Finally, in addition to the errors summarized above, the covariance
matrix is further modified due to systematic uncertainties in the
purity and completeness of the sample.  The basic set up is this: if
$N_{true}(N_{200})$ is the number of clusters one expects in the absence
of systematics, and $N_{obs}(N_{200})$ is the actual observed number of
clusters, one has
\begin{equation}
N_{obs}(N_{200}) = \lambda(N_{200}) N_{true}(N_{200}),
\end{equation}
where $\lambda$ is a factor close to unity that characterizes the purity and completeness systematics.  
If the sample is pure but incomplete, $\lambda$ is simply equal to the sample's completeness.  For
a complete but impure sample, $\lambda$ is one over the sample's purity.   Note that, in general,
$\lambda$ is itself a function of the cluster richness $N_{200}$.
In \citet{rozoetal07a}, we estimated the purity and completeness of the maxBCG
cluster sample at $95\%$ or higher for $N_{200} \ge 11$ (see Figures 3 and 6 in that paper),
suggesting $\lambda=1.00\pm 0.05$. 
Given that 
\begin{equation}
\Var(N_{obs}) = \Var(\lambda)N_{true}^2 + \lambda^2\Var(N_{true}),
\end{equation}
it follows that we can incorporate the impact of this nuisance
parameter by simply adding in quadrature the relative uncertainty
introduced by $\lambda$ to the covariance matrix estimated in the
previous section.  
A similar argument holds for the total mass contained in clusters
within each richness bin.  That is, if $\bar M_{true}$ is the true
mean mass of clusters of richness $N_{200}$, and $\bar M_{obs}$ is the
observed mean mass, we expect
\begin{equation}
(N \bar M)_{obs} = \tilde \lambda (N\bar M)_{true},
\end{equation}
where $\tilde\lambda$ is a correction factor that accounts for the
mass contribution of impurities in the sample.  Unfortunately, it is
impossible to know a priori what this factor $\tilde\lambda$ should
be, {\it even if we knew the correction factor $\lambda$
 for cluster abundances}.  The reason is that false cluster detections
 will most certainly have a mass overdensity associated with them,
 just not that of a halo of the expected mass given the observed richness.
 Without a priori knowledge of this mass contribution, it is impossible
 to estimate the proper value of $\tilde \lambda$.
In the extreme case that
all false detections have mass $\bar M$,
then the recovered value for $N\bar M$ will be
biased by a factor $\tilde\lambda =
\lambda$, which suggests adopting a fiducial value
$\Var(\tilde\lambda)=0.05^2$ to add to the diagonal matrix elements
corresponding to the observed weak lensing masses.  That is the approach
we follow here.  Throughout, we always set $\Var(\tilde\lambda)=\Var(\lambda)$.


\subsection{Model Parameters and Priors}
\label{sec:priors}

Our analysis assumes a neutrino-less, flat $\Lambda$CDM cosmology, and we
fit for the values of $\sigma_8$ and $\Omega_m$.  The Hubble parameter
is held fixed at $h=0.7$, and the tilt of the primordial power
spectrum is set to $n=0.96$ as per the latest WMAP results
\citep{wmap08}.  The baryon density $\Omega_bh^2$ is also held fixed
at its WMAP5 value $\Omega_bh^2=0.02273$.  Of these secondary
parameters, the two that are most important are the Hubble constant
and tilt of the primordial matter power spectrum \citep{rozoetal04}.
Section \ref{sec:results_cosmology} demonstrates our results are
robust to marginalization over these additional parameters.

The richness--mass relation $P(N_{200}|M)$ is assumed to be a log-normal
of constant scatter. The mean log-richness at a given mass $\avg{\ln
  N_{200}|M}$ is assumed to vary linearly with mass, resulting in two
free parameters.  We comment on possible deviations from linearity in
Section \ref{sec:3pmean}.  For the two parameters specifying the mean
richness--mass relation we have chosen the value of $\avg{\ln N_{200}|M}$
at $M=1.3\times 10^{14}\ \msun$ and at $M=1.3\times 10^{15}\ \msun$.
These two are very nearly the values of the mean mass for our lowest
and highest richness bins, and therefore roughly bracket the range of
masses probed in our analysis.  The value of $\avg{\ln N_{200}|M}$ at any
other mass is computed through linear interpolation.  We adopt flat
priors on both of these parameters.

The scatter in the richness--mass relation $\sigma_{N_{200}|M}$ is
defined as the standard deviation of $\ln N_{200}$ at fixed $M$,
$\sigma_{N_{200}|M}^2 = \Var(\ln N_{200}|M)$.  We assume that this quantity
is a constant that does not scale with mass, and adopt a flat prior
$\sigma_{N_{200}|M}\in[0.1,1.5]$ for this parameter.  We comment on
possible deviations from constant scatter in Section
\ref{sec:2pscatt}.  The minimum scatter allowed in our work
($\sigma_{N_{200}|M}=0.1$) corresponds to a $10\%$ scatter, which is the
predicted scatter for $Y_X$ in simulations.  $Y_X$ is usually regarded
as the X-ray mass tracer that is most tightly correlated with mass, so
our prior on the scatter is simply the statement that richness
estimates are less faithful mass tracers than $Y_X$.

We also
place a prior on the converse scatter, that is, the scatter in mass at fixed richness
$\sigma_{M|N_{200}}^2 = \Var(\ln M|N_{200})$ at $N_{200}=40$. 
We emphasize that in our analysis the scatter $\sigma_{M|N_{200}}$ 
is considered an observable,
not a parameter (the parameters is $\sigma_{N_{200}|M}$).  The probability distribution
 $P(\sigma_{M|N_{200}})$ is taken directly from the analysis
by \citet{rozoetal08a}, and can be roughly summarized as
$\sigma_{M|N_{200}}=0.45\pm0.10\ (1\ \sigma)$.  This constraint is
derived by demanding consistency between the observed $\Lx-N_{200}$
relation of maxBCG clusters, the mass--richness relation of maxBCG
clusters derived from weak lensing, and the $\Lx-M$ relation of
clusters measured in the 400d survey \citep{vikhlininetal08b}.
To compute the observed scatter $\sigma_{M|N_{200}}$ as a function
of our model parameters 
we directly compute the variance in log-mass for clusters in a richness
bin $N_{200} \in [38,42]$.  The variance in $\ln M$ due to the finite
width of the bin is of order $(1/40)^2 \approx 0.006$, which is to be
compared to the intrinsic variance $\approx 0.45^2 \approx 0.2$.
Because the intrinsic variance is significantly larger than the
variance due to using a finite bin width, our results are not
sensitive to the width of the bin used in the implementation of the
prior.  We have explicitly checked that this is indeed the case. We
have also checked that our results are insensitive to the location of
the richness bin.  That is, placing our prior on $\sigma_{M|N_{200}}$ at
$N_{200}=30$ and $N_{200}=50$ gives results that are nearly identical to
those obtained with our fiducial $N_{200}=40$ value.
Finally, we note that in using the scatter measurement of
\citet{rozoetal08a}, who used an overdensity threshold of 500 relative
to critical to define cluster masses, we are making the implicit
assumption that the value of the current uncertainties in the scatter
are much larger than any sensitivity to differences in the cluster
mass definition.  To address this concern, in Section
\ref{sec:results_scatter} we discuss how the scatter prior
impacts our results.

The redshift selection function $P(z_{photo}|z)$ is assumed to be
Gaussian with $\avg{z_{photo}|z}=z$ and $\sigma(z_{photo}|z)=0.008$,
as per the discussion in \citet{koesteretal07a}.  We have explicitly
checked that our results are not sensitive to our choice of parameters
within the range $\delta\avg{z_{photo}|z}\approx 0.005$ and
$\delta\sigma_{z_{photo}|z}=0.02$, which encompass the uncertainties
in the photometric redshift distribution of the maxBCG clusters
\citep{koesteretal07a}.

Finally, we also adopt a prior on the weak lensing mass bias parameter,
$\beta=1.0\pm0.06$, and allow it to vary over the range
$[0.5,1.5]$.  The width of our Gaussian prior is simply the mean
difference between the \citet{johnstonetal07} masses (after correcting
for photometric redshift bias) and those of \citet{mandelbaumetal08b}
\citep[for a more detailed discussion see][]{rozoetal08a}.  

The total number of parameters that are allowed to vary in our Monte Carlo Markov Chain (MCMC) 
is six: $\sigma_8$, $\Omega_m$, $\avg{\ln N_{200}|M}$
evaluated at $M=1.3\times 10^{14}\ \msun$ and $M=1.3\times 10^{15}\ \msun$, $\sigma_{N_{200}|M}$, and $\beta$.  We summarize
the relevant priors in Table \ref{tab:priors}.


\begin{deluxetable}{ccc}
\tablewidth{0pt}
\tablecaption{\label{tab:priors}}
\tablehead{Parameter$^{a}$ & Prior$^{b}$ & Importance$^{c}$}
\startdata
$\sigma_8$ & [0.4,1.2] & \ unrestrictive\  \\
$\Omega_m$ & [0.05,0.95] &  \ unrestrictive\ \\
$\avg{\ln N_{200}|M_1}$ & flat &  \ unrestrictive\ \\
$\avg{\ln N_{200}|M_2}$ & flat &  \ unrestrictive\ \\
$\sigma_{N_{200}|M}$ & [0.1,1.5] &  \ unrestrictive\ \\
$\beta$ & $1.00\pm 0.06$; $[0.5,1.5]$ & restrictive \\
\hline
$\sigma_{M|N_{200}}\ ^{d}$ & \citet{rozoetal08a} & restrictive  \\
\enddata
\tablenotetext{a}{The masses $M_1$ and $M_2$ are set to $1.3\times
  10^{14}\ \msun$ and $1.3\times 10^{15}\ \msun$ respectively.}
\tablenotetext{b}{Priors of the form $[a,b]$ mean the parameter in
  question is restricted to values within that range (flat prior).
  Priors of the form $x=a\pm \delta a$ refer to a Gaussian prior of
  mean $\avg{x}=a$ and variance $\Var(x)=(\delta a)^2$.  }
\tablenotetext{c}{Column specifies whether our results are sensitive
  to the assumed priors.  We refer to a prior as restrictive if our
  cosmological constraints are sensitive to the assumed prior, and
  unrestrictive otherwise.  The only restrictive priors are that on
  the mass bias parameter $\beta$ and the prior on the scatter in mass
  at fixed richness.}  
  \tablenotetext{d}{Note $\sigma_{M|N_{200}}$ is not really a parameter
  in our analysis but an observable that can be computed given the six
  parameters above.}
\end{deluxetable}



\subsection{Implementation}

We use the low baryon transfer functions of \citet{eisensteinhu99} to
estimate the linear matter power spectrum.  The halo mass function is
computed using \citet{tinkeretal08}.  We use a mass definition
corresponding to a 200 overdensity with respect to the mean matter
density of the universe, and adopt the Sheth-Tormen expressions for
the mass dependence of halo bias \citep{shethtormen02} (this enters into
our analysis only in the calculation of sample variance).  The
likelihood function is sampled using a Monte Carlo Markov Chain (MCMC)
approach with a burn in of 22,000 points during which the covariance
matrix of the parameters is continually updated so as to provide an
ideal sampling rate \citep{dunkleyetal05}.  We then run the chains for
$10^5$ points, and use the resulting outputs to estimate the $68\%$
and $95\%$ likelihood contours in parameter space.  For further
details, we refer the reader to \citet{rozoetal07b}.

The one point that is worth discussing here is our corrections for
the dependence of the recovered weak lensing masses on the assumptions
about cosmology used for the measurements.  \citet{johnstonetal07}
quote halo masses at an overdensity of 180 relative to the mean
background of the universe.  Given that we use a density contrast of
200 relative to mean in order to compute the halo mass function, we
must re-scale the observed masses to our adopted mass definition.
Moreover, the weak lensing analysis assumed $\Omega_m=0.27$.  Given a
different matter density parameter $\tilde \Omega_m$, the quoted mass
will no longer correspond to an overdensity of $180$, but to an
overdensity of $180(0.27/\tilde \Omega_m)$.  We explicitly apply this
re-scaling to the observed weak lensing masses at each point in our
MCMC.  In practice, there is also an additional correction due to the
dependence of the lensing critical surface density $\Sigma_c$ on the
matter density parameter $\Omega_m$, as well as small corrections due
to systematic variations in halo concentration with mass.  However,
these corrections are expected to be small, and are fully degenerate
with the mass bias parameter $\beta$, so we do not include them here.
The rescaling of the weak lensing masses is done using the fitting
formulae in \citet{hukravtsov03}.


\section{Results}
\label{sec:results}

\begin{figure*}[t]
\scalebox{0.87}{\rotatebox{90}{\plotone{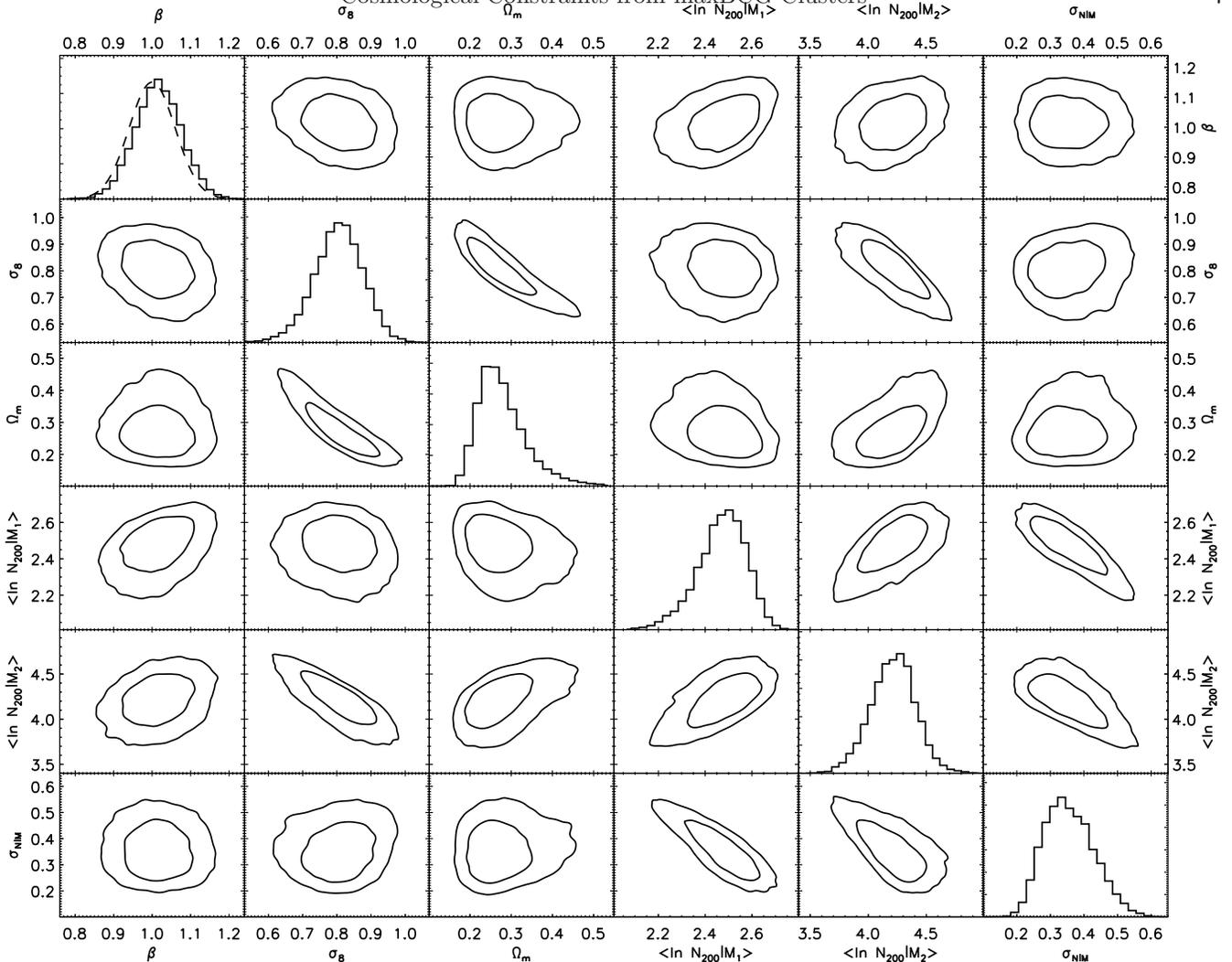}}}
\caption{ Confidence regions for each pair of parameters that were
  allowed to vary in our fiducial analysis (described in \S
  \ref{sec:analysis}).  Contours show $68\%$ and $95\%$ confidence
  regions.  Plots along the diagonal show the probability
  distributions for each quantity marginalized over the remaining
  parameters.  The probability distribution for the mass bias
  parameter $\beta$ also shows the prior $\beta=1.00\pm0.06$ assumed
  in the analysis.}
\label{fig:Tableplot}
\end{figure*}

Figure~\ref{fig:Tableplot} presents the $68\%$ and $95\%$ confidence
regions for each pair of parameters in our fiducial analysis described
in \S \ref{sec:analysis}.  Plots along the diagonal show the
probability distributions of each quantity marginalized over the
remaining parameters.  Upper left plot showing the probability
distribution of the mass parameter $\beta$ also shows the prior
$\beta=1.00\pm 0.06$ as a dashed curve. Our best fit model is summarized in Table 
\ref{tab:bestfit}, and is defined as the expectation value of all of our parameters.
To test that our best fit model is a good model to the data, we performed
$10^4$ Monte Carlo realizations of our best fit model, and evaluated the likelihood
function for each of these realizations.  Setting $\avg{\ln \lk}=0$, from our Monte Carlo
realizations we find $\ln \lk = 0.0\pm 6.9$, which is to be compared to the data likelihood
$\ln \lk=-5.2$.  The data likelihood is therefore consistent with our model, demonstrating
the model is statistically a good fit.


In the discussion
that follows, we restrict ourselves to the subset of plots which
we find most interesting.   Throughout, unless otherwise noted we summarize
constraints on a parameter $p$ by writing $p=\bar p + \sigma_p$ where
$\bar p$ and $\sigma_p$ are the mean and standard deviation of the
likelihood distribution for $p$ marginalized over all other
parameters.  We use this convention even when the likelihood function
is obviously not Gaussian.  


\begin{deluxetable}{ccc}
\tablewidth{0pt}
\tablecaption{\label{tab:bestfit} Best Fit Model}
\tablehead{Parameter$^{a}$ & maxBCG & maxBCG+WMAP5$^b$}
\startdata
$\sigma_8$ & $0.804\pm0.073$  & $0.807\pm 0.020$\\
$\Omega_m$ & $0.281\pm0.066$  &  $0.269\pm0.018$ \\
$\avg{\ln N_{200}|M_1}$ & $2.47\pm0.10$  &  $2.48\pm0.10$ \\
$\avg{\ln N_{200}|M_2}$ & $4.21\pm0.19$ &  $4.21\pm0.13$ \\
$\sigma_{N_{200}|M}$ & $0.357\pm0.073$ & $0.348\pm0.071$ \\
$\beta$ & $1.016\pm0.060$ & $1.013\pm0.059$ 
\enddata
\tablenotetext{a}{The masses $M_1$ and $M_2$ are set to $1.3\times
  10^{14}\ \msun$ and $1.3\times 10^{15}\ \msun$ respectively.}
\tablenotetext{b}{These values are obtained by including the WAMP5
prior $\sigma_8(\Omega_m/0.25)^{-0.312}=0.790\pm0.024$.  See
Section \ref{sec:degeneracies} for details.}
\end{deluxetable}


\subsection{Cosmological Constraints and Comparison to WMAP}

\begin{figure}[t!]
\epsscale{1.2}
\plotone{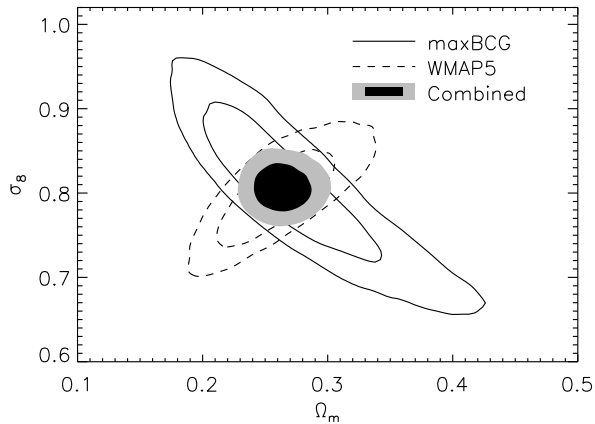}
\caption{
Constraints on the $\sigma_8-\Omega_m$ plane from maxBCG and
 WMAP5 for a flat $\LCDM$ cosmology.  Contours show the $68\%$ and
 $95\%$ confidence regions for maxBCG (solid), WMAP5 (dashed), and
  the combined results (filled ellipses).  The thin axis of the
  maxBCG-only ellipse corresponds to
 $\sigma_8(\Omega_m/0.25)^{0.41}=0.832\pm 0.033$.  The joint
 constraints are $\sigma_8=0.807\pm 0.020$ and $\Omega_m=0.265\pm
 0.016$ (one-sigma errors).
}
\label{fig:s8_Om}
\end{figure}

The solid curves in Figure~\ref{fig:s8_Om} show the $68\%$ and $95\%$
confidence regions from our analysis.  The ``thin'' axis of our error
ellipse corresponds to $\sigma_8(\Omega_m/0.25)^{0.41}=0.832\pm
0.033$.\footnote{The exponent $0.41$ is obtained by estimating the covariance
matrix of $\ln \sigma_8$ and $\ln \Omega_m$, and finding the best constrained
eigenvector.}  The constraints on each of the individual parameters are
$\sigma_8=0.80\pm0.07$ and $\Omega_m=0.28\pm0.07$.  The marginalized
likelihood can be reasonably approximated by a log-normal distribution with
$\ln \Omega_m=-1.313\pm 0.183 $, $\avg{\ln \sigma_8}=-0.219 \pm 0.081$,
and a correlation coefficient between $\ln \Omega_m$ and $\ln \sigma_8$
$r=-0.899$.
Also shown in Figure~\ref{fig:s8_Om} as
dashed curves are the corresponding regions from the WMAP 5-year
results \citep{wmap08}.  Our results are consistent with WMAP5.
Combining the two experiments results in the inner filled ellipses,
given by $\sigma_8=0.807\pm0.020$ and $\Omega_m=0.265\pm0.016$, with
nearly no covariance between the two parameters ($r=0.008$).  These
joint constraints on $\sigma_8$ and $\Omega_m$ represent nearly a
factor of two improvement relative to the constraints from WMAP alone.



\begin{figure}[t!]
\epsscale{1.2}
\plotone{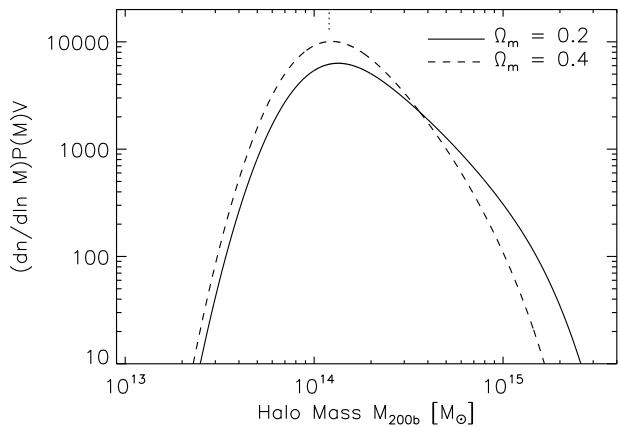}
\caption{Halo mass function for two different cosmologies satisfying
  the maxBCG constraint $\sigma_8(\Omega_m/0.25)^{0.41}=0.832$.  The
  mass functions are weighted by the volume probed by the maxBCG
  catalog (computed assuming $\Omega_m=0.265$), and by the mass
  selection function shown in Figure~\ref{fig:binning}.  The maxBCG
  normalization condition $\sigma_8(\Omega_m/0.25)^{0.41}=0.832$
  results in a fixed halo abundance at a mass scale $M=3.6\times
  10^{14}\ \msun$.  The dotted line at the top marks the mass scale at
  which the mean of the richness--mass relation is best constrained in
  our fiducial analysis.}
\label{fig:mf}
\end{figure}

The shape of the confidence region is easy to interpret:
since the number of
massive clusters increases with both $\sigma_8$ and $\Omega_m$, in
order to hold the cluster abundance fixed at its observed value any
increase in $\sigma_8$ must be compensated by a decrease in
$\Omega_m$, implying that a product of the form
$\sigma_8\Omega_m^\gamma$ must be held fixed.  The specific value of
$\gamma$ depends on the mass scale that is best constrained from the
data.  The particular degeneracy recovered by our analysis corresponds
to a mass scale $M=3.6\times 10^{14}\ \msun$, which is about what we
would expect (i.e. roughly half way between the lowest and highest
masses probed by our data).  Figure~\ref{fig:mf} illustrates this
argument by showing the \citet{tinkeretal08} halo mass function
weighted by the mass selection function from Figure~\ref{fig:binning}
for two different cosmologies: a low $\sigma_8$ (high $\Omega_m$)
cosmology, and a high $\sigma_8$ (low $\Omega_m$) cosmology, where the
product $\sigma_8\Omega_m^{0.41}$ has been held fixed to our best-fit
value.  We will refer back to Figure~\ref{fig:mf} multiple times
in the following discussion.


\subsection{Constraints on the Richness--Mass Relation}

In our analysis, we parameterized the richness--mass relation in terms
of its scatter, and the value of the mean $\avg{\ln N_{200}|M}$ at two
mass scales, $M_1=1.3\times10^{14}\ \msun$ and $M_2=1.3\times
10^{15}\ \msun$.  We now re-parameterize this relation in terms of an
amplitude and slope for $\avg{\ln N_{200}|M}$, selecting as the pivot
point the mass scale at which the uncertainty in $\avg{\ln N_{200}|M}$ is
minimized.  We write then
\begin{equation}
\avg{\ln N_{200}|M} = A + \alpha( \ln M - \ln M_{pivot} ).
\label{eq:ampdef}
\end{equation}
We find the error on the amplitude parameter is minimized for
$M_{pivot}=1.09\times 10^{14}$, which agrees well with the peak in the
mass distribution of our clusters as shown in Figure~\ref{fig:mf}.  In
what follows, we discuss only constraints on the richness--mass
relation assuming this parameterization.  A discussion of possible
curvature in the richness--mass relation and/or mass scaling of its
scatter is relegated to \S \ref{sec:parsys}.

\begin{figure}[t!]
\epsscale{1.15}
\plotone{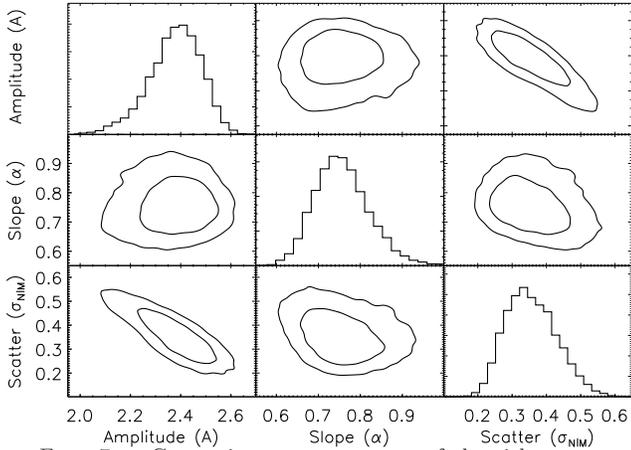}
\caption{Constraints on parameters of the richness--mass relation.
  Countours indicate the $68\%$ and $95\%$ confidence regions;
  diagonal histograms indicate the probability distribution for each
  parameter, marginalized over the remaining parameters.  The
  amplitude and slope parameters define the mean of the richness--mass
  relation as per Eq. \ref{eq:ampdef}.  The pivot point of the
  relation occurs at $M=1.15\times 10^{14}\ \msun$.}
\label{fig:nu-M}
\end{figure}

Figure~\ref{fig:nu-M} summarizes our constraints on the richness--mass
relation after marginalizing over all other parameters.  The best-fit
values for each of the parameters are $A \def \avg{\ln
  N_{200}|M_{pivot}}=2.34\pm 0.10$, $\alpha=0.757\pm0.066$, and
$\sigma_{N_{200}|M}=0.357\pm 0.073$.  Note that for a pure power-law
abundance function, one expects 
$\sigma_{N_{200}|M} =\alpha \sigma_{M|N_{200}}$, in accordance with
our result.

Of these results, the constraints on the slope and scatter of the
richness--mass relation are particularly worth noting.  First, it is
clear that the naive scaling $N_{200}\propto M$ is not satisfied, with
the slope of the richness--mass relation being significantly smaller
than unity.  Second, the recovered scatter $\sigma_{N_{200}|M}=0.357\pm
0.073$ is larger than the Poisson value $\sigma_{N_{200}|M}\approx 0.2$
that one might naively expect for clusters with $N_{200}\approx 30$
galaxies, which is the typical richness of clusters at the mass scale
where mass function is best constrained.

Interpreting these results in terms of standard halo occupation model
parameters requires care.  The maxBCG
richness is known to suffer from various sources of systematics
including miscentering of clusters \citep{johnstonetal07b} and color
off-sets in the richness estimates \citep{rozoetal08b}, both of which
will impact the recovered richness--mass relation at some level.
Moreover, {\it any} richness estimate will suffer to some extent from
projection effects \citep{cohnetal07}, and discrepancies between
assigned cluster radii and the standard mass-overdensity definitions
used for halos.  Disentangling the various contributions of each of
these different sources of scatter to the total variance of the
richness--mass relation is beyond the scope of this paper, and will
not be considered further here.


Figure~\ref{fig:nu-M} also shows that the amplitude of the
richness--mass relation is anti-correlated with the scatter.  This is
not surprising: at fixed cluster abundance, and given a fixed mass
function, models with a high amplitude of the richness--mass relation
result in halos that tend to be
very rich.  This means that the number of lower mass halos that
scatter into higher richness must be low, or otherwise the abundance
of clusters will be over-predicted.  Consequently, high amplitude
models must have low scatter, leading to an anti-correlation between
the two parameters.


\subsection{Degeneracies Between Cosmology and the Richness--Mass Relation}
\label{sec:degeneracies}

Figure~\ref{fig:Tableplot} shows that the most significant correlation
between cosmology and our fiducial richness--mass relation parameters
is that between $\sigma_8$ and $\avg{\ln N_{200}|M_2}$ where $M_2$ is our
higher reference mass $M_2=1.3\times 10^{15}\ \msun$.  Because the
pivot point for the mean of the richness--mass relation is so close to
our original low mass reference scale $M_1=1.3\times
10^{14}\ \msun$ used to define $\avg{\ln N_{200}|M}$, it follows that $M_2$
must be closely related to $\alpha$, the slope of the richness--mass
relation.   We thus expect a strong degeneracy between $\sigma_8$ and
$\alpha$ \citep[see also][]{rozoetal04}.

\begin{figure}[t]
\epsscale{1.1}
\plotone{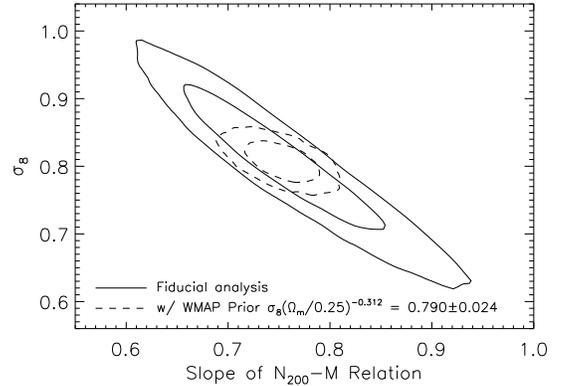}
\caption{Confidence regions in the $\sigma_8-\alpha$ plane.  Solid
  ellipses show the $68\%$ and $95\%$ likelihood regions.  The tight
  correlation between $\sigma_8$ and $\alpha$, the slope of the
  richness--mass relatio, can be understood on the basis of
  Figure~\ref{fig:mf}: a low $\sigma_8$ implies few massive halos, so
  to avoid under-predicting the abundance of rich clusters, galaxies
  must preferentially live in lower mass halos, resulting in a more
  rapidly rising richness--mass relation (i.e. higher slope). This
  degeneracy is broken upon inclusion of the WMAP five-year constraint
  $\sigma_8(\Omega_m/0.25)^{-0.312}=0.790\pm0.024$ as an additional
  prior, as illustrated by the inner dashed ellipses in the Figure.
  The corresponding constraint on the slope of the richness--mass
  relation is $\alpha=0.752\pm0.024$.}
\label{fig:s8-slope}
\end{figure}

Figure~\ref{fig:s8-slope} shows that this is indeed the case.  We can
understand the origin of this anti-correlation by investigating
Figure~\ref{fig:mf}.  We have seen that the data fixes the amplitude
of the halo abundance at $M=3.9\times 10^{14}\ \msun$.  At the high
mass end, however, the expected abundance of massive halos varies
rapidly with $\sigma_8$.  Low $\sigma_8$ models result in fewer
massive halos, so high richness clusters will have relatively lower
masses.  That is, richness must increase steeply with mass, and hence
$\alpha$ must be high, explaining the anti-correlation between
$\sigma_8$ and $\alpha$.

Figure~\ref{fig:s8-slope} also demonstrates how these constraints are
improved when we include a WMAP five-year data prior
$\sigma_8(\Omega_m/0.25)^{-0.312}=0.790\pm0.024$.  This prior
corresponds to the error along the thin direction of the WMAP error
ellipse.  Since WMAP data breaks the $\sigma_8-\Omega_m$ degeneracy in
the data, including the WMAP prior produces a tight constraint in the
$\sigma_8-\alpha$ plane.  The new marginalized uncertainty in the
slope of the richness--mass relation is $\alpha=0.752\pm 0.024$,
significantly smaller than unity.



\section{Systematic Errors}
We now consider the impact of three varieties of systematic errors on
our analysis. Section \ref{sec:observational_systematics}
investigates observational systematics, Section
\ref{sec:prior_systematics} investigates systematics due to our
assumed priors, and Section \ref{sec:parsys} investigates
systematics due to the parameterization of the richness--mass
relation.

\subsection{Observational Systematics}
\label{sec:observational_systematics}

In this section, we study how observational systematics affect the
recovered cosmological constraints from our analysis.  We consider two
such systematics: one, the impact of purity and completeness, and
two, the impact of possible biases in the weak lensing mass
estimates of the maxBCG clusters.  We do not discuss uncertainties in
the photometric redshifts for clusters at any length since, as
discussed in Section \ref{sec:priors}, they are found to be
negligible.  This is not surprising, as the maxBCG photometric
redshift estimates are extremely accurate \citep[$\sigma_z\approx 0.008$,][]{koesteretal07a}.

\subsubsection{The Impact of Purity and Completeness}
\label{sec:results_purcomp}

\begin{figure}[t]
\epsscale{1.2}
\plotone{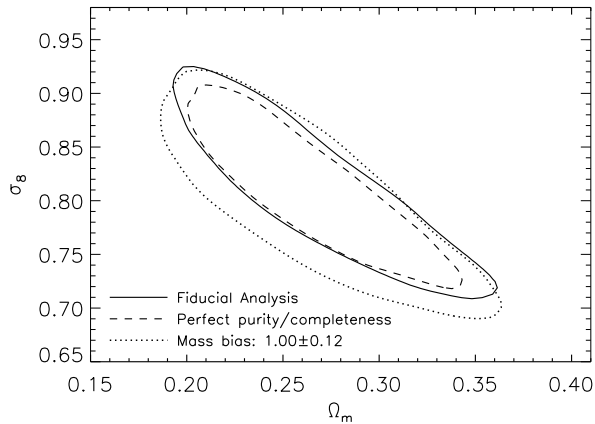}
\caption{Effect of purity, completeness, and mass bias on parameter constraints.
  Plot shows $68\%$ confidence regions assuming perfect purity and
  completeness (dashed), increasing the width of the weak lensing mass
  bias prior from $\beta=1.00\pm 0.05$ to $\beta=1.00\pm 0.12$
  (dotted), and for our fiducial analysis (solid).  
  We find the uncertainty in purity and completeness has a
  minimal impact on the best constrained combination of the $\sigma_8$
  and $\Omega_m$ parameter, and therefore on the constraints from a
  joint maxBCG + WMAP5 analysis.  The same is not true of the weak
  lensing mass bias parameter.  Uncertainties in the maxBCG cluster
  masses are the dominant source of systematic in our current
  analysis, and increase the uncertainty of the parameter combination
  $\sigma_8(\Omega_m/0.25)^{0.4}$ by $45\%$ (see Figures
  \ref{fig:prior_sys} and \ref{fig:par_sys} for comparison).  }
\label{fig:obs_sys}
\end{figure}

Figure~\ref{fig:obs_sys} compares the cosmological constraints
obtained assuming perfect purity and completeness with those obtained
assuming a $5\%$ uncertainty in these quantities.  While
non-negligible, the $5\%$ uncertainty in the completeness and purity
function of the maxBCG catalog is far from the dominant source of
uncertainty in our analysis.  Moreover, this uncertainty elongates the
error ellipse along its unconstrained direction, but has a minimal
impact on the best constrained combination of $\sigma_8$ and
$\Omega_m$: $\Delta \sigma_8(\Omega_m/0.25)^{0.41} = 0.033$ in our
fiducial analysis, while $\Delta \sigma_8(\Omega_m/0.25)^{0.41}
=0.029$ assuming perfect purity and completeness, a mere $10\%$
difference.

It is easy to understand why a $5\%$ uncertainty in the purity and
completeness has a minimal impact in our results.  For $N_{200}\gtrsim
25$, the statistical uncertainties in the cluster abundances are
larger than the $5\%$ uncertainty in the counts from purity and
completeness.  Since the best constrained combination of cosmological
parameters is driven primarily by high mass clusters, a $5\%$
uncertainty in the purity and completeness functions has little impact
on this parameter combination.  How far the error ellipse extends
along the degeneracy, however, is primarily driven by the
observational constraints on the low end of the halo mass function
(see Figure~\ref{fig:mf}).  Consequently, the $5\%$ systematic
uncertainty in the low richness cluster counts elongates the error
ellipse along its major axis.

We conclude that for the expected level of purity and completeness of
the maxBCG cluster sample, our cosmological constraints are robust to
these systematics.


\subsubsection{Systematic Uncertainties of the Weak Lensing Mass Estimates}
\label{sec:results_WLsys}

In Section \ref{sec:masses}, we discussed that the weak lensing masses
of \citet{johnstonetal07} were boosted by a factor of $1.18$ to
account for biases arising from scatter in the photometric redshift
estimates \citep{mandelbaumetal08}.  Even with such a boost, the
\citet{johnstonetal07} and the \citet{mandelbaumetal08b} mass
estimates were not consistent, which led us in Section
\ref{sec:expectation_values} to introduce a mass bias parameter
$\beta$ that uniformly scales all masses by the same amount in order
to account for any remaining biases.  We now wish to explore how
robust our results are to our estimate of this systematic uncertainty.

Figure~\ref{fig:obs_sys} illustrates what happens if we repeat
our fiducial analysis while doubling the width of the prior of $\beta$
from $\beta$ from $\beta=1.00\pm 0.06$ to $\beta=1.00\pm0.12$.  We
find that the wider $\beta$ prior significantly increases the
uncertainty in the parameter combination $\sigma_8(\Omega_m/0.25)$
from $\Delta\sigma_8(\Omega_m/0.25)^{0.41}=0.033$ to
$\Delta\sigma_8(\Omega_m/0.25)^{0.41}=0.045$, corresponding to a
$36\%$ increase of the error bar.  Using this new, wider prior, we
find that the joint maxBCG + WMAP 5-year likelihood result in the
cosmological constraints $\sigma_8=0.802\pm0.023$ and
$\Omega_m=0.261\pm0.019$, which constitute a $\approx 15\%$ increase
in the uncertainty of each of these parameters respectively.  Even
with this wider prior, however, adding the maxBCG constraint to the
WMAP5 result improves the final cosmological constraints on $\sigma_8$
and $\Omega_m$ by a factor of 1.6 relative to those obtained using
WMAP data alone.

We can understand the impact of the mass bias parameter on our
cosmological constraints using Figure~\ref{fig:mf}.  A wider prior on
$\beta$ implies that the mass scale of the maxBCG clusters is more
uncertain, so the mass at which the cluster abundance is best
constrained, i.e.  the point at which the two curves in
Figure~\ref{fig:mf} cross each other, is more uncertain.
Consequently, the cluster normalization constraint
$\sigma_8\Omega_m^{0.41}$ is weakened.  The error along the long
direction of the error ellipse does not change because the width of
the mass range probed by the maxBCG clusters is largely independent of
an overall mass bias.

One of the curious results that we have found in our study of the mass
bias parameter $\beta$ is that the prior and posterior distributions
of this parameter are different.  In particular, we find that given
the priors $\beta=1.00\pm 0.06$ and $\beta=1.00\pm 0.12$, the
posterior distributions for $\beta$ are $\beta=1.02\pm 0.06$ and
$\beta=1.06\pm 0.12$ respectively. Indeed, this explains why the error
ellipse for our wider prior is displaced to the left of that of our
fiducial analysis: the shift in $\beta$ corresponds to a change in the
mass scale, which has to be compensated by a change in the matter
density parameter $\Omega_m$.

We conclude that the uncertainty in the weak lensing mass estimates of
the maxBCG clusters is an important source of systematic uncertainty
in our analysis.  In fact, it is the dominant source of systematic
uncertainty in our analysis.  We have explicitly considered the impact
of photometric redshift estimates for source galaxies as the source of
this uncertainty, but other biases to the lensing masses --- for
example if the fraction of miscentered clusters was over- or
under-estimated by \cite{johnstonetal07} --- would affect our results
in a similar way.


\subsection{Prior-Driven Systematics}
\label{sec:prior_systematics}

Our analysis makes use of two important priors:  that the only
two cosmological parameters of interest are $\sigma_8$ and $\Omega_m$,
and that the scatter in the richness--mass relation can be determined
from X-ray studies as discussed in \citet{rozoetal08a}.
Here, we discuss how our results change if these priors are relaxed.

\subsubsection{Cosmological Priors}
\label{sec:results_cosmology}

\begin{figure}[t]
\epsscale{1.2}
\plotone{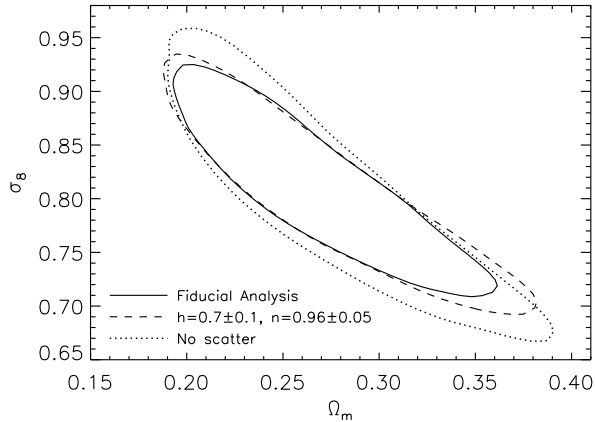}
\caption{ Effect of relaxing additional cosmological parameter priors
  on $\sigma_8$ and $\Omega_m$ constraints.  Lines show $68\%$
  confidence regions for our fiducial analysis (solid), after
  marginalizing over Gaussian priors $h=0.7\pm 0.1$ and $n=0.96\pm
  0.05$ (dashed), and using only a flat prior on the scatter in mass
  at fixed richness $\sigma_{M|N_{200}}\in[0.1,1.5]$ (dotted).  We find
  that holding the Hubble parameter and power spectrum index fixed
  does not bias nor artificially tighten our constraint on
  $\sigma_8\Omega_m^{0.41}$.  The scatter prior from
  \citet{rozoetal08a} on $\sigma_{M|N_{200}}$ employed in our fiducial
  analysis is found to have a significant impact on our data.  More
  specifically, dropping this scatter prior increases the error along
  the short axis of our error ellipses by $36\%$.  We have also
  explored whether massive neutrinos significantly impact our
  constraint on $\sigma_8\Omega_m^{0.41}$, and find that for neutrino
  masses $\sum m_\nu < 1\ \mbox{eV}$ there is no degradation of the
  error.  }
\label{fig:prior_sys}
\end{figure}

After $\sigma_8$ and $\Omega_m$, cluster abundance studies are most
sensitive to the Hubble parameter $h$ and the tilt $n$ of the
primordial power spectrum.  In Figure~\ref{fig:prior_sys}, we
illustrate how the constraints on the $\sigma_8-\Omega_m$ plane are affected
upon marginalization over $h$ and $n$ using Gaussian priors $h=0.7\pm
0.1$ and $n=0.96\pm 0.05$. As we can see, marginalizing over the
Hubble parameter and the tilt of the power spectrum elongates the
error ellipse, but it does not make it wider.  Thus, the combination
$\sigma_8\Omega_m^{0.41}$ remains tightly constrained, and a joint
maxBCG and WMAP 5-year data analysis is robust to the details of the
priors used for $h$ and $n$ when estimating the maxBCG likelihood
function.  We also investigated whether a non-zero neutrino mass could
significantly affect our results.  Using a prior $\sum m_\nu< 1\
\mbox{eV}$, we find that massive neutrinos do not significantly affect
our constrain on $\sigma_8(\Omega_m/0.25)^{0.41}$.  We conclude that
holding the Hubble parameter and the tilt of the power spectrum fixed
does not result in systematic uncertainties in the joint maxBCG + WMAP
5-year data analysis.


\subsubsection{The Impact of the Scatter Prior}
\label{sec:results_scatter}

In \citet{rozoetal08a}, we derived an empirical constraint on the
scatter of the richness--mass relation by demanding consistency
between X-ray, weak lensing, and cluster abundance data.  The
recovered scatter, however, characterized the richness--mass relation
using a mass that was defined using an overdensity of 500 relative to
the critical density of the universe.  In this analysis, we use a
density threshold of 200 relative to mean, so the use of the X-ray
derived scatter prior is justified only if the scatter in the mass
scaling between the two overdensity thresholds is not the dominant
source of scatter.  While we fully expect this assumption to hold, we
have repeated our analysis without use of the scatter prior in order
to cross-check our results.

Figure~\ref{fig:prior_sys} summarizes our results.  We find that our
scatter prior tightens the error ellipse along both its short and long
axis.  This is as expected: without the scatter prior, the mass scale
of the maxBCG clusters becomes less constrained, and consequently the
halo mass function is less tightly constrained at all scales.  The
best constrained combination of $\sigma_8$ and $\Omega_m$ when
dropping the \citet{rozoetal08a} prior on the scatter in the
mass--richness relation is $\sigma_8(\Omega_m/0.25)^{0.48} =
0.841\pm0.045$.  This value represents a $36\%$ increase in
uncertainty relative to our fiducial analysis.  The joint maxBCG +
WMAP5 constraints in this case are $\sigma_8=0.805\pm0.021$ and
$\Omega_m=0.264\pm0.017$.

Not surprisingly, prior knowledge of the scatter of the mass--richness
relation can significantly enhance the constraining power of the
maxBCG data set.  Nevertheless, even without prior knowledge in the
scatter the joint maxBCG+WMAP constraints improve upon the WMAP values
by a factor of 1.7.


\subsection{Parameterization Systematics}
\label{sec:parsys}

One of the most important systematics that need to be addressed in studies where
the observable--mass relation is parameterized in some simple way is how to
assess the robustness of the results to changes in the parameterization of the
observable--mass relation.  Here, we have assumed that the richness--mass relation
$P(N_{200}|M)$ is a log-normal of constant scatter and that $\avg{\ln
  N_{200}|M}$ varies linearly with $\ln M$.  We now investigate how our
results change if we relax some of these assumptions.

\subsubsection{Curvature in the Mean Richness--Mass Relation}
\label{sec:3pmean}

To investigate the impact of curvature in the mass richness relation,
we assume $\avg{\ln N_{200}|M}$ is a piecewise linear function.  We first
specify $\avg{\ln N_{200}|M}$ at three mass scales $M_1$, $M_2$, and
$M_3$, and define the value of $\avg{\ln N_{200}|M}$ at every other mass
through linear interpolation in log-space.  We set the minimum and
maximum reference masses to the same values as before, $M_1=1.3\times
10^{14}\ \msun$, and $M_3=1.3\times 10^{14}\ \msun$.  The intermediate
reference mass is set to the geometric average of these two masses,
$\ln M_2 = 0.5(\ln M_1+\ln M_3)$, or $M_2=3.66\times 10^{14}\ \msun$.
Note this mass scale is very nearly the same as the mass at which the
halo mass function is best constrained.

\begin{figure}[t]
\epsscale{1.2}
\plotone{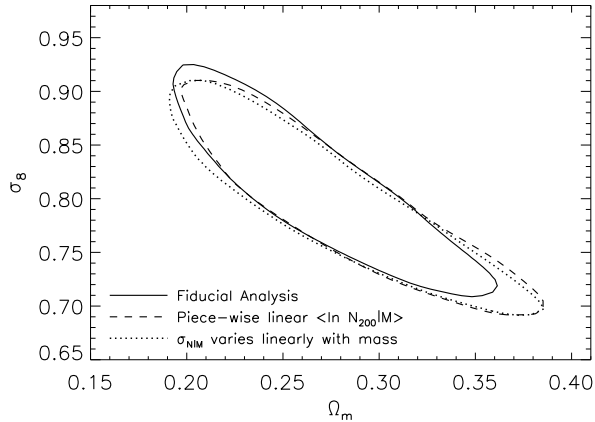}
\caption{Effect of relaxing assumptions about the richness--mass
  relation on $\sigma_8$--$\Omega_m$ constraints.  Contours show
  $68\%$ confidence limits for our fiducial analysis (solid curve), 
  assuming $\avg{\ln N_{200}|M}$
  is a piece-wise linear function (dashed), and allowing
  $\sigma_{N_{200}|M}$ to vary linearly with mass (dotted).  Giving
  additional freedom to the richness--mass relation has a minimal
  impact on our constraint on $\sigma_8\Omega_m^{0.41}$.  Moreover,
  using a likelihood ratio test we find that there is no evidence in
  the data for curvature of the richness--mass relation, nor for a
  scatter that varies with mass.  We conclude that our
  parameterization of $\avg{\ln N_{200}|M}$ and $\sigma_{N_{200}|M}$ do not
  introduce any significant systematics in our analysis.  }
\label{fig:par_sys}
\end{figure}

Figure~\ref{fig:par_sys} shows how our cosmological constraints change
with the introduction of mass dependence on the slope of the mean
richness--mass relation $\avg{\ln N_{200}|M}$.  We find that the thin
axis of the error ellipse is not significantly affected by this more
flexible parameterization, while the long axis of the error ellipse is
somewhat lengthened.  This is as expected: the high mass end of the
halo mass function is only sensitive to how richness varies with mass for
large $M$, and in this regime the more flexible parameterization does
not introduce significantly more freedom.  Thus, our data will tightly
constrain the high mass end of the halo mass function just as well as
did before, leading to no degradation in the error of
$\sigma_8\Omega_m^{0.41}$.  Once the high mass end of the
richness--mass relation has been fixed, however, introducing curvature
in $\avg{\ln N_{200}|M}$ dilutes the information contained in the low
mass end of the halo mass function, thereby increasing the error
ellipse along its long axis.  Note the robustness of the
$\sigma_8\Omega_m^{0.41}$ constraint also implies that the constraints
of a joint maxBCG + WMAP5 analysis are not significantly affected by
our choice of parameterization.


Irrespective of the impact our new parameterization of $\avg{\ln
  N_{200}|M}$ has on our cosmological constraints, it is fair to ask
whether or not there is significant evidence for curvature of the mean
richness--mass relation.  Using a maximum likelihood ratio test, we
find that the increase in likelihood due to curvature in the
richness--mass relation is significant at the $50\%$ level, less than
$1\sigma$.  Thus, there is no evidence for curvature in the
richness--mass relation.  We have also explicitly confirmed that the
slopes of the low and high mass end of the richness--mass relation are
consistent with each other.  Indeed, we find
\begin{equation}
\left . \frac{d^2\avg{\ln N_{200}|M}}{d \ln M^2} \right \vert_{M=3.66\times 10^{14}\ \msun}= 0.05\pm 0.07
\end{equation}
where we have assumed
\begin{equation}
\frac{d^2\avg{\ln N_{200}|M}}{d \ln M^2} =
	\frac{f(M_3)+f(M_1)-2f(M_2)}{ 0.5^2(\ln M_3 - \ln M_1)^2}.
\end{equation}
and $f(M)=\avg{\ln N_{200}|M}$.


\subsubsection{Scaling of the Scatter in the Richness--Mass Relation with Mass}
\label{sec:2pscatt}

We now investigate whether allowing the scatter of the richness--mass
relation to vary with mass has a significant impact on our
cosmological parameters.  For these purposes, we allow the scatter
$\sigma_{N_{200}|M}$ to vary linearly with $\ln M$, and parameterize it
by specifying its values at the reference masses $M_1=1.3\times
10^{14}\ \msun$ and $M_2=1.3\times 10^{15}\ \msun$.  The value of
$\sigma_{N_{200}|M}$ at any other mass is obtained through linear
interpolation.

Figure~\ref{fig:par_sys} compares the cosmological constraints we
obtain with our new model to those of our fiducial analysis with
constant scatter.  Once again, we find that the ``thin'' axis of the
error ellipse is not significantly affected by the new more flexible
parameterization, while the long axis is slightly elongated.  The
interpretation of these results is the same as those of \S
\ref{sec:3pmean}.  We have tested for evidence of scaling of the
scatter in the richness--mass relation with halo mass using a
likelihood ratio test.  The increase in likelihood due to a linearly
varying scatter is significant at the $39\%$ level, implying there is
no evidence of mass dependence in the scatter of the richness--mass
relation in the data.  We have also explicitly confirmed that the
scatter at the low and high mass ends probed by the maxBCG cluster
sample are consistent with each other.  Indeed, our constraint on the
slope of the mass dependence of the scatter in the richness--mass
relation is
\begin{equation}
\left . \frac{d\sigma_{N_{200}|M}}{d \ln M} \right \vert_{M=3.66\times 10^{14}\ \msun} = 0.00\pm0.06
\end{equation}
where we assumed
\begin{equation}
\frac{d\sigma_{N_{200}|M}}{d \ln M} = \frac{ \sigma_{N_{200}|M_2}-\sigma_{N_{200}|M_1} }{\ln M_2 - \ln M_1}.
\end{equation}
We note the velocity dispersion analysis in
\citet[see][]{beckeretal07} points towards some mass dependence in the
scatter of the mass--richness relation, though part of this
discrepancy is likely due to miscentering systematics \citep[see][for
details]{rozoetal08a}.  We are now in the process of reanalyzing the
velocity dispersion data updating both our treatment of systematics,
and substantially increasing the sample of spectroscopically sampled
galaxies, so we defer a detailed discussion of these results to a
future paper.

We conclude that our parameterization of the mean and scatter of the richness--mass relation does not introduce systematic
errors in our analysis.


\subsubsection{Richness Range Considered}
\label{sec:richrange}

\begin{figure}[t]
\epsscale{1.2}
\plotone{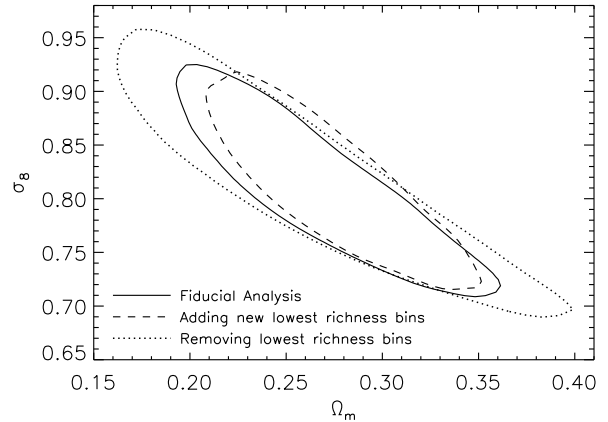}
\caption{Effect of the minimum richness on $\sigma_8$--$\Omega_m$
  constraints.  Coutours show $68\%$ confidence regions for our
  fiducial analysis (solid curve), an analysis where we include an
  additional richness bin, $\nu = 9-10$, at the low end of the
  richness function (dashed), and an analysis where we drop the lowest
  richness bins considered in our fiducial analysis.  Our cosmological
  constraints are consistent for all these analyses.  We have also
  found that removing the most massive clusters from our analysis has
  minimal impact on our cosmological constraints.  }
\label{fig:richness_range}
\end{figure}

We have tested whether there is cosmological information in the
richness range $N_{200}>120$ by running MCMCs both with and without
the contribution of these clusters to the likelihood function.  We
find that these two analyses yield nearly identical results.  We have
also explicitly confirmed that our results are robust to the lowest
richness bin employed in the analysis.  As we might expect, removing
the lowest richness bin increases our uncertainties along the long
axis of the error ellipse as shown in Figure~\ref{fig:richness_range}.
We also investigate adding a new lowest richness bin, consisting of
clusters in with $N_{200} = $ 9--10, as well as the mean mass for
clusters in the range $N_{200} = $ 9--11.  This analysis rotates the
error ellipse very slightly compared to our fiducial analysis, but
does not significantly affect our results.


\section{Discussion}
\subsection{Comparison to Other Work}
\label{sec:comparison}

The main point of this section is to demonstrate two points: 
\begin{enumerate}
\item The cosmological constraints from the maxBCG cluster catalog are
  competitive with the state of the art constraints derived from low
  redshift X-ray selected cluster samples.
\item Despite the markedly different analyses and sources of
  systematic uncertainty, the cluster abundance constraints from the
  maxBCG cluster sample are in excellent agreement with those of X-ray
  selected samples.  This demonstrates the robustness of cluster
  abundance studies as a tool of precision cosmology.
\end{enumerate}

Given our goal, in this section we focus exclusively on the most
recent cosmological constraints derived from low redshift X-ray
cluster samples.  In particular, we explicitly consider only three
works: \cite{mantzetal08}, who worked with the X-ray luminosity
function, \cite{henryetal08}, who worked with the X-ray temperature
function, and \cite{vikhlininetal08b}, who estimated the low redshift
halo mass function using the 400d X-ray survey \citep{bureninetal07}
with mass estimates based on $Y_X$ \citep{kravtsovetal06}.  
These three papers are the most
recent analyses of X-ray selected cluster samples, and all recover
tight cosmological constraints that are in excellent agreement with
one another, while carefully accounting for the relevant systematics
for each of their analyses.

Now, as we have discussed in previous sections, the main result from
low redshift cluster abundance studies is a tight constraint on the
value of $\sigma_8\Omega_m^\gamma$ where for maxBCG clusters
$\gamma=0.41$.  Other cluster samples, however, will have slightly
different values of $\gamma$, which brings up the question of how can
we fairly compare these various constraints.  One way would be to
simply quote the percent uncertainty in the relevant
$\sigma_8\Omega_m^\gamma$ combination.  However, we would like to have
a clear graphical representation of this result.  We have chosen to do
this by plotting the $68\%$ confidence regions of a simplified version
of a joint cluster abundance + WMAP5 analysis assuming a neutrino-less
flat $\Lambda$CDM cosmology.  We proceed as follows: given a cluster
abundance experiment, we consider only the constraint on
$\sigma_8\Omega_m^\gamma$, disregarding all other cosmological
information.  We then add a WMAP 5 prior
$\sigma_8(\Omega_m/0.25)^{-0.312}=0.790\pm0.024$, which corresponds to
the thin axis of the error WMAP5 error ellipse in the
$\sigma_8-\Omega_m$ plane, and we compute the corresponding $68\%$
confidence regions in the $\sigma_8-\Omega_m$ plane.


\begin{figure}[t]
\epsscale{1.2}
\plotone{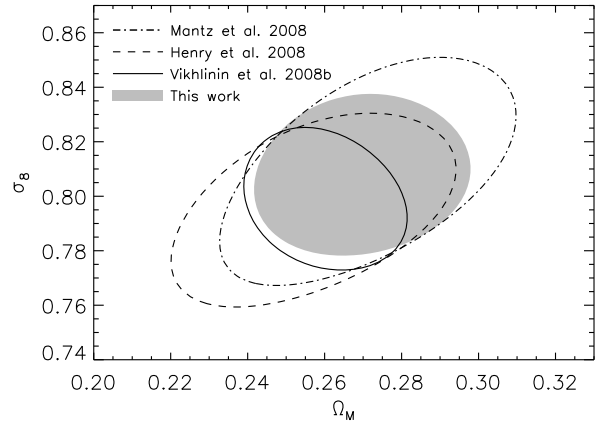}
\caption{Comparion of optical and X-ray cluster abundance constraints
  on $\sigma_8-\Omega_m$.  Contours show $68\%$ confidence regions
  for a joint WMAP5 and cluster abundance analysis assuming a flat
  $\Lambda$CDM cosmology.  In addition to our results (filled ellipse), we
  consider the latest cluster abundance constraints from the low
  redshift cluster luminosity function \citep[dashed][]{mantzetal08},
  temperature function \citep[dash-dot][]{henryetal08}, and mass
  function as estimated with $Y_X$ \citep[solid][]{vikhlininetal08b}.
  All four studies are in excellent agreement with each other despite
  the tight cosmological constraints and the different sources of
  systematic uncertainty among the various analyses.  }
\label{fig:compare_abundance_constraints}
\end{figure} 



\begin{deluxetable*}{cccccc}
  \tablewidth{0pt} \tablecaption{\label{tab:abundance_constraints}
    Cosmological Constraints From Multiple Cluster Abundance
    Experiments} \startdata
\tablehead{Source & Reference & $\gamma$ & $\sigma_8
  (\Omega_m/0.25)^\gamma$ & $\sigma_8$ & $ \Omega_m$}
\startdata
maxBCG Richness Function & this work & 0.41 & $0.832\pm 0.033$ & $0.807\pm 0.020$ & $0.270\pm 0.019$ \\
X-ray Luminosity Function & \citet{mantzetal08} & 0.62 & $0.85\pm 0.07$ & $0.809\pm 0.028$ & $0.272\pm 0.026$ \\
Temperature Function & \citet{henryetal08} & 0.30 & $0.80 \pm 0.04$ & $0.795\pm 0.023$ & $0.258\pm 0.025$ \\
Mass function estimated with $Y_X $ & \citet{vikhlininetal08b} & 0.47 & $0.808\pm 0.024$ & $0.798\pm 0.017$ & $0.260\pm 0.014$ \\
\enddata
\tablecomments{
The $\sigma_8$ and $\Omega_m$ constraint from the
  maxBCG + WMAP5 analysis quoted here differs very slightly from that
  presented in Figure~\ref{fig:s8_Om} because of the simplified
  approach we have taken in this section for deriving the constraints
  (see text for details).  Note \citet{vikhlininetal08b} quote their
  result as $\sigma_8 (\Omega_m/0.25)^{0.47}=0.813\pm 0.013\ (stat)\
  \pm 0.02\ (sys)$.  For this study, we have simply added these two
  uncertainties in quadrature. }
\end{deluxetable*}


The result of this exercise is shown in
Figure~\ref{fig:compare_abundance_constraints}.  The specific
constraints from each of the works considered here are presented in
Table~\ref{tab:abundance_constraints}.  The agreement among the
different analyses is excellent despite the tight error bars and the
different sources of systematic uncertainties.  This agreement clearly
demonstrates not only that optically selected cluster samples can
produce cosmological constraints that are competitive with those of
X-ray selected cluster samples, but also that systematic uncertainties
have been properly estimated.


\subsection{Low Redshift Cluster Abundances and The Equation of State of Dark Energy}
\label{sec:w}

Detailed analyses exploring how cluster abundances help improve dark
energy constraints have been presented by previous groups, most
recently by \citet{mantzetal08} and \citet{vikhlininetal08b}.  Rather
than duplicating their work, in this section we opt for performing a
simple analysis that captures the essential physics behind the
\citet{mantzetal08} and \citet{vikhlininetal08b} results, which helps
illustrate exactly why and how clusters complement 
CMB, supernova, and BAO studies.


\begin{figure*}[t]
\plottwo{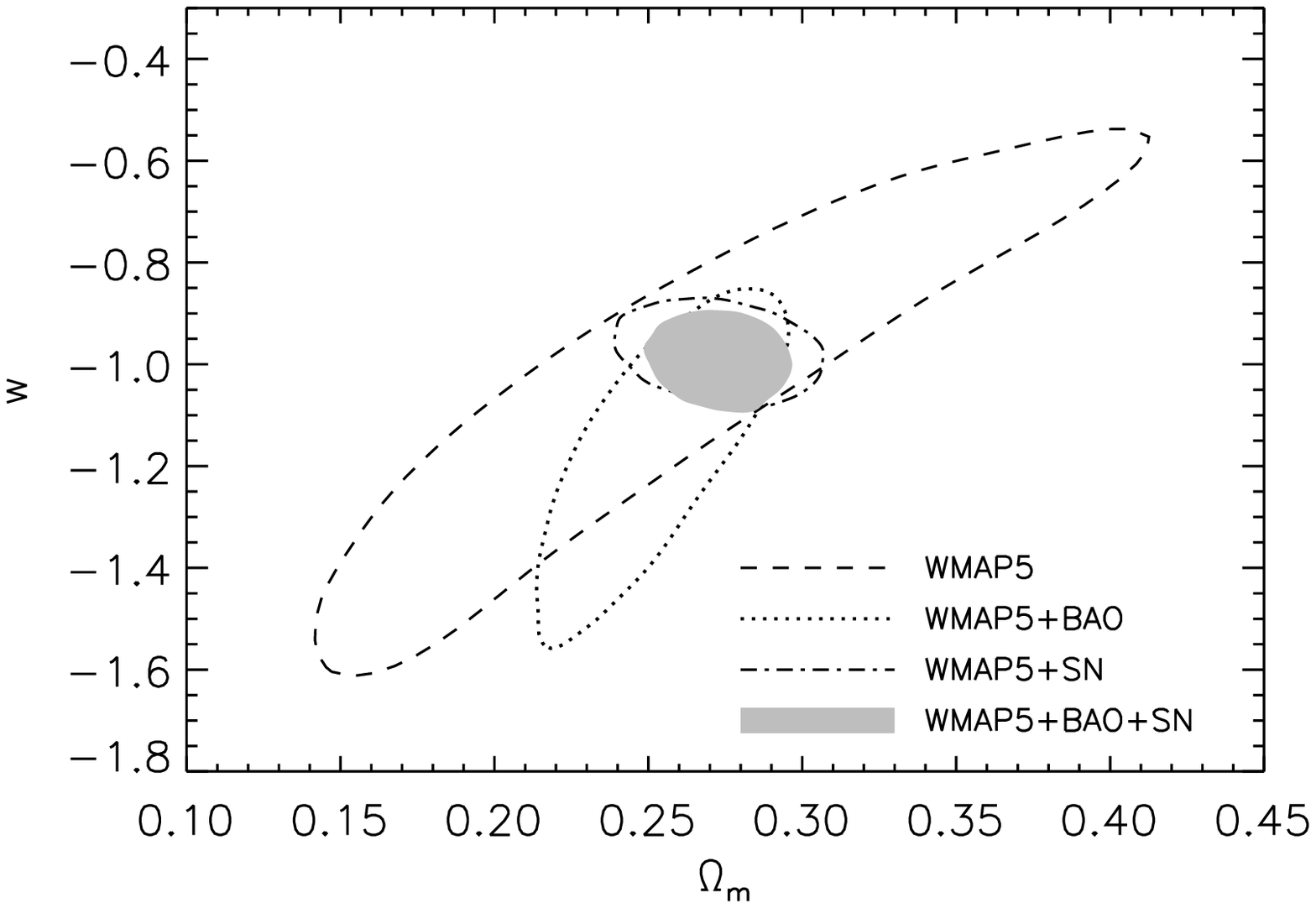}{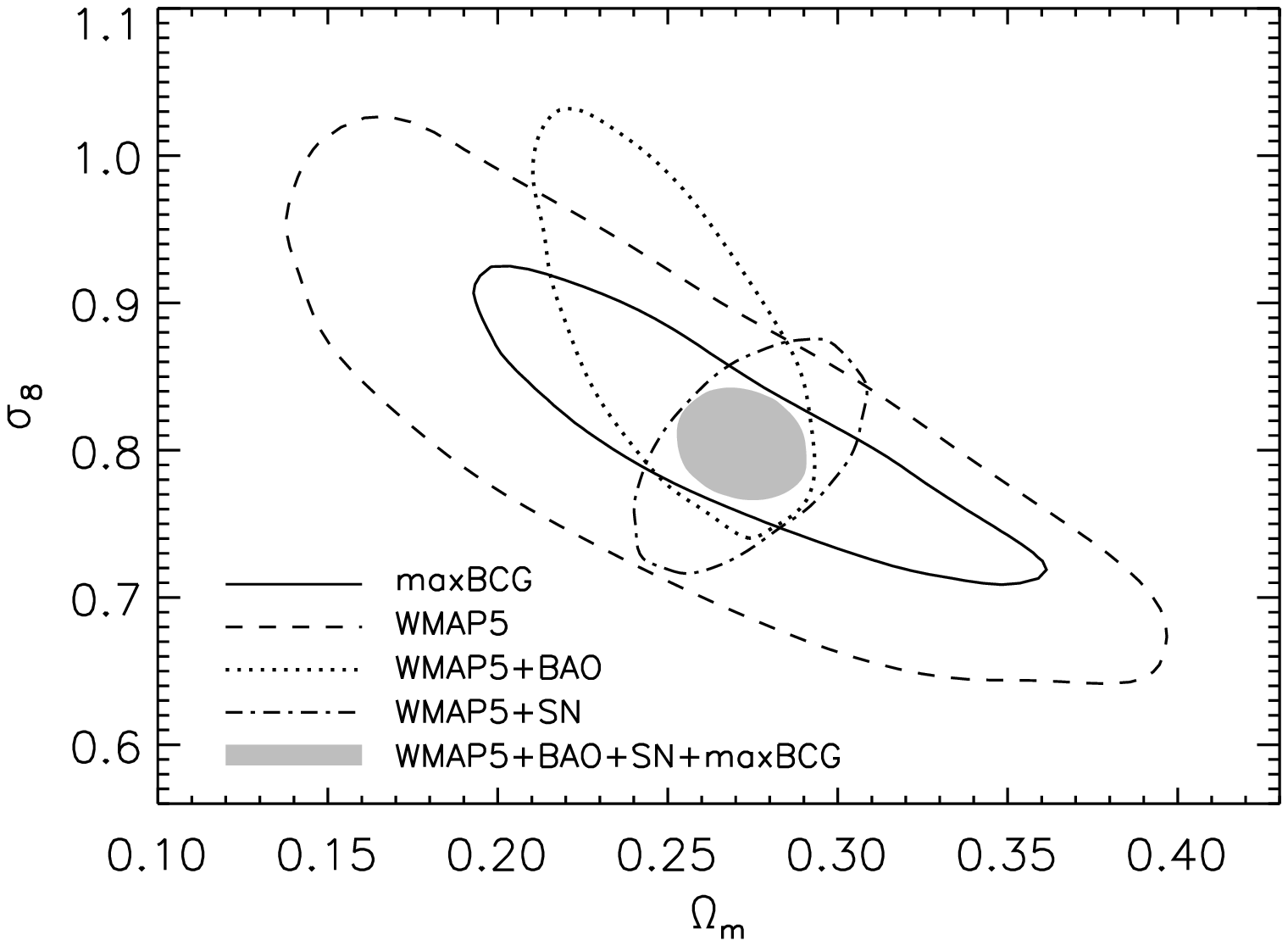}
\caption{Parameter constraints on the $w-\Omega_m$ plane (left)
  and $\sigma_8-\Omega_m$ plane (right)
  in a flat $w$CDM cosmology, for various data combinations.  All contours
  shown are $68\%$ confidence, and are obtained using the MCMC chain
  outputs downloaded from the LAMBDA website
  (http://lambda.gsfc.nasa.gov/).  Despite the fact that the WMAP5
  data constrain the amplitude of the primordial power spectrum with
  comparable accuracy in both a $\Lambda$CDM and $w$CDM cosmology,
  allowing $w$ to vary introduces a large degeneracy between $w$ and
  $\Omega_m$.  This degeneracy severely degrades the WMAP constraints
  in the $\sigma_8-\Omega_m$ plane, as seen in the right panel.
  Adding new observables that break the $w-\Omega_m$ degeneracy
  restores the complementarity between WMAP5 and clusters in the
  $\sigma_8-\Omega_m$ plane, which helps improve dark energy
  constraints through the growth of structure. }
\label{fig:evolDE1}
\end{figure*} 

 
We begin by focusing on the somewhat surprising result by
\citet{vikhlininetal08b} that a joint WMAP5 and low redshift cluster
abundance experiment does {\it not} produce an interesting constraint
on the equation of state of dark energy $w$.  The reason this is
surprising is that WMAP5 has measured the amplitude of the power
spectrum at recombination to high accuracy.  Given this value and a
cosmological model, one can predict the value of $\sigma_8$ today.  By
demanding that this prediction agrees with the cluster normalization
condition, one ought to obtain a tight constraint on the dark energy
equation of state.

To understand why this is not the case, consider first the WMAP5
results.  The parameters $w$ and $\Omega_m$ are
strongly degenerate given the WMAP5 data alone, as shown in
Figure~\ref{fig:evolDE1}.  The value of $\sigma_8$ implied by the
WMAP5 data depends sensitively on these two parameters, so a large
uncertainty in $w$ and $\Omega_m$ dramatically increases the area of
the $\sigma_8-\Omega_m$ plane allowed by the WMAP5 data.  Moreover, we
can see from Figure~\ref{fig:evolDE1} that the WMAP constraint goes
from being orthogonal to the cluster normalization condition to being
parallel to it, implying that the cluster normalization condition cannot
improve upon the dark energy constraints of WMAP alone.  Indeed, a
prior of the form $\sigma_8(\Omega_m/0.25)^{0.41}=0.832\pm0.033$ has a
minimal impact on the error bar in $w$.  Fortunately, given this
understanding, it is easy to see how to improve this situation: we
need to introduce an additional observable which breaks the
$w-\Omega_m$ degeneracy.  As an example, in the above Figures we also
show the $68\%$ confidence intervals obtained for three additional
analyses:
\begin{enumerate}
\item A joint WMAP5+BAO analysis, which includes the dark energy
  constraints derived by \citet{eisensteinetal05} using the Baryon
  Acoustic Oscillation (BAO) measurement from the SDSS LRG galaxy
  sample.
\item A joint WMAP5+SN analysis, which draws on the Union 
combined dataset \citep{kowalskietal08}, a compilation
of Super Nova (SN) data composed of the 
  \citet[][gold sample only]{riessetal04}, \citet{astieretal06}, and
  \citet{miknaitisetal07} supernova samples.
\item A joint WMAP5+BAO+SN analysis, which adds both BAO and SN
  measurements as extra observables.
\end{enumerate}
In all cases, the confidence contours are estimated based on the MCMC
data made publicly available by the WMAP team through the LAMBDA
website (http://lambda.gsfc.nasa.gov/).  These data sets break the
$w-\Omega_m$ degeneracy from the WMAP5 data alone, and they restore the
complementarity between WMAP5 and the cluster normalization condition
in the $\sigma_8-\Omega_m$ plane, as illustrated in the lower panel of
Figure~\ref{fig:evolDE1}.


\begin{figure}[t]
\epsscale{1.2}
\plotone{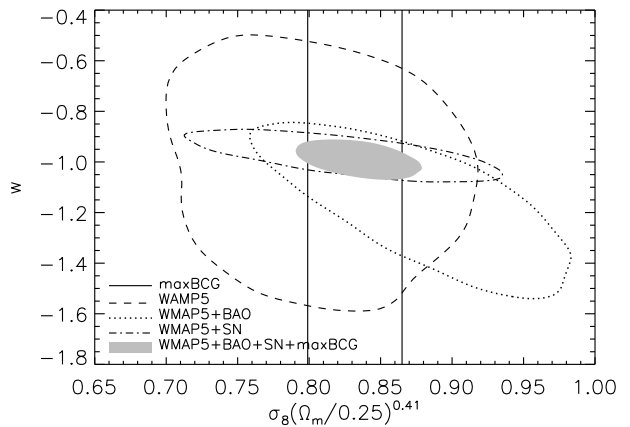}
\caption{Confidence contours in the $w-\sigma_8(\Omega_m/0.25)^{0.41}$
  plane for analyses using various combinations of cosmological data.
  Contours all indicate 68\% confidence.  Provided the $w-\Omega_m$
  degeneracy from Figure~\ref{fig:evolDE1} is broken by an additional
  observable, cluster abundances can help constrain dark energy
  through the growth of structure between the time of last scattering
  and the low-redshift universe.  }
\label{fig:w-e1}
\end{figure} 

 

\begin{deluxetable*}{cccc}
\tablewidth{0pt}
\tablecaption{\label{tab:DE} Cosmological Constraints in a Flat $w$CDM Cosmology}
\tablehead{Experiment & $\Omega_m$  & $w$ & $\sigma_8$ }
\startdata
WMAP5	 &  $0.266\pm0.086$ & $-1.05\pm0.34$ & $0.811\pm 0.121$ \\
WMAP5+BAO & $0.251\pm0.027$ & $-1.20\pm0.24$ & $0.885\pm0.094$ \\
WMAP5+SN & $0.274\pm0.023$ & $-0.98\pm0.07$ & $0.798\pm 0.053$ \\
WMAP5+maxBCG & $0.265\pm0.048$ & $-1.07\pm 0.34$ & $0.815\pm 0.061$ \\
WMAP5+SN+BAO & $0.274\pm 0.015$	& $-0.995\pm0.067$ & $0.808\pm0.047$ \\
WMAP5+SN+maxBCG & $0.274\pm0.016$ & $-0.978\pm 0.053$ & $0.801\pm 0.026$ \\
WMAP5+BAO+maxBCG & $0.258\pm0.023$ & $-1.097\pm 0.160$ & $0.831\pm0.044$\\
WMAP5+BAO+SN+maxBCG & $0.272\pm0.013$ & $-0.989\pm0.053$ & $0.805\pm0.026$ \\
\enddata
\tablecomments{
  The constraints quoted here are derived by multiplying the WMAP5
  likelihoods with a Gaussian prior of our cluster normalization
  condition.  We demonstrate in the text that results derived in this
  way are nearly identical to those from more detailed treatments.  }
\end{deluxetable*}


Figure~\ref{fig:w-e1} shows the constraints in the
$w-\sigma_8(\Omega_m/0.25)^{0.41}$ plane for the various analyses
considered in Figure~\ref{fig:evolDE1}.  The corresponding
cosmological constraints are summarized in Table \ref{tab:DE}.  In
order to compute how cluster abundances improve cosmological
constraints, we have simply added the cluster normalization condition
derived within the standard $\Lambda$CDM cosmological model as a
prior.  While
one might worry that letting the dark energy equation of state vary
would degrade the uncertainty in the cluster normalization
condition, in practice $w$ has a minimal impact provided one
restricts oneself to low redshift cluster samples.
For instance, varying $w$ by $\Delta w=0.1$ changes the comoving distance
to the median redshift of maxBCG clusters by $\approx 1\%$.  The
growth function is even less sensitive, varying by a mere $\approx
0.3\%$.  Thus, the cluster normalization condition from low-redshift
cluster samples is essentially independent of $w$.

To demonstrate this explicitly, we compare the results of
\citet{vikhlininetal08b} to the results from our simple analysis in
which the cluster normalization condition
$\sigma_8(\Omega_m/0.25)^{0.47}=0.808\pm 0.024$ from
\citet{vikhlininetal08b} is added to the data sets mentioned above.
For a joint WMAP+BAO+clusters analysis, they find $w=-0.97\pm0.12$
while our simple analysis results in $w=-1.04\pm0.13$.  Note that the error
bars are nearly identical, while the central value for $w$ differ by
only half a $\sigma$.  This is still true after adding supernovae as
an additional constraint, in which case they find $w=-0.991\pm0.045$
compared to our $w=-0.971\pm0.048$.   A similar conclusion can be
reached for the \citet{mantzetal08} analysis.  For a joint WMAP+supernovae
+$f_{gas}$+cluster abundance analysis, the find $w=-1.02\pm0.06$.  With
our simple analysis, and ignoring $f_{gas}$, we obtain $w=-0.98\pm0.06$.
This demonstrates that, to high
accuracy, current cluster catalogs improve cosmological constraints on
dark energy only through the low-redshift cluster normalization
condition.

In summary, we have shown that cluster abundances help constrain the
dark energy equation of state principally through the cluster
normalization condition at low redshifts, which constrains the growth
of structure between the epoch of recombination and today.  However,
the intrinsic degeneracy between $w$ and $\Omega_m$ given CMB data
renders this test ineffective unless the degeneracy is broken by an
additional data set.  It is also worth remarking here that if we
compare the results of a WMAP5+SN analysis to those obtained after
including the maxBCG cluster normalization condition, the constraint
on the dark energy equation of state is only improved at the $25\%$
level, going from $\Delta w = 0.07$ to $\Delta w=0.054$.  This
reflects the fact that distance-redshift relationships tend to be more
sensitive to $w$ than the growth of structure.  Nevertheless, the good
agreement between the WMAP5+SN constraints and the cluster
normalization is far from trivial.  Indeed, the WMAP5+SN likelihood
contours in the $\sigma_8-\Omega_m$ plane assume general relativity,
so the good agreement with our data indicates that we are not able to
resolve any departures from Einstein's theory of gravity \citep[see
also][]{rapettietal08,mortonsonetal09}.  While quantitative constraints
on such deviations are model dependent, our final error on $\sigma_8$
allow us to unambiguously state that
models for which the growth factor between last scattering and today
differ from our best fit $\Lambda$CDM model by $\approx 6\%$ can
be ruled out at the $2\sigma$ level.


\subsection{Prospects For Improvement}
\label{sec:improvement}

It is worth considering to what extent we can expect the cosmological
constraints from maxBCG to improve with further study.  Given that the
two principal sources of systematic uncertainty are the amplitude of
the weak lensing mass calibration and the prior on the scatter of the
mass--richness relation, we focus here on those two quantities.  More
specifically, we re-analyze our data using artificially tight priors
on each of these parameters individually, as well as on both
parameters simultaneously.  The tight priors adopted for this exercise
are $\beta=1.00\pm0.01$ for the mass bias parameter,
and $\sigma_{M|N_{200}}=0.45\pm 0.02$ for the scatter in mass at fixed
richness.  We find that the uncertainties in $S_8 =
\sigma_8(\Omega_m/0.25)^{0.41}$ for each of these analyses are:
\begin{enumerate}
\item fiducial analysis: $\Delta S_8=0.033$,
\item tight scatter prior alone: $\Delta S_8=0.026$,
\item tight mass bias prior alone: $\Delta S_8=0.025$, and
\item tight mass bias and scatter prior: $\Delta S_8=0.018$.
\end{enumerate}
Thus, a tight prior on either the mass bias or the scatter parameter
improves the principal maxBCG cosmological constraint by $25\%$. If
both priors are tightened, the improvement is as high as $50\%$.  The
corresponding constraint on $w$ for this most optimistic scenario,
assuming a joint WMAP5+BAO+SN+maxBCG analysis, would be $\Delta
w=0.049$, which is only a $7\%$ improvement relative to the current
constraint.

Are such improvements feasible?  In principle, yes.  Improvement of
the mass bias parameter is possible through a follow-up spectroscopic
program aimed at calibrating the mean lensing critical surface density
of the lens-source pairs used to estimate the mean cluster masses.
Likewise, an extensive X-ray follow-up program could in principle
constrain the scatter in the mass--richness relation to high accuracy.
In practice, realizing such tight priors might be difficult.  For
instance, given the current scatter estimate
$\sigma_{M|N_{200}}=0.45$, we require $\approx 400$ X-ray follow ups
to achieve an uncertainty of $\Delta \sigma_{M|N_{200}} \approx 0.02$.
Such an extensive program seems unlikely to be feasible any time in
the near future.  What is needed, then is a way to significantly
reduce the number of follow-up observations necessary to improve our
cosmological constraints.  Fortunately, the number of follow-up
observations necessary to achieve such accuracy scales as the square
of the scatter $\sigma_{M|N_{200}}$, so the best thing to do at this
point is probably to focus on constructing new richness estimators
that better correlate with halo mass \citep[see
e.g.][]{rozoetal08b,reyesetal08}.

Finally, we note that there remains additional information about the maxBCG
clusters that has not yet been incorporated into our analysis.  This includes
galaxy velocity dispersion data \citep{beckeretal07}, X-ray data 
\citep[][while X-ray data was used to place a constraint on the scatter in mass at
fixed richness, we did not otherwise use the X-ray data in this analysis]{rykoffetal08a}, 
and clustering
information \citep{estradaetal08}.  Including these additional probes of cluster
mass should help further improve our cosmological constraints.


\section{Summary}
\label{sec:summary}

We have performed a joint analysis of the abundance and weak lensing mass
estimates of the maxBCG clusters detected using SDSS imaging data.  In addition
to this data,  a prior on the scatter in the mass--richness relation
derived from demanding consistency between the weak lensing and X-ray
mass estimates of the clusters.  Our cosmological constraints can be
summarized as
\begin{equation}
\sigma_8(\Omega_m/0.25)^{0.41}=0.832\pm0.033,
\end{equation}
which is
consistent with and complementary to the latest WMAP results.  With a joint maxBCG and
WMAP5 analysis we find
\begin{eqnarray}
\sigma_8 & = & 0.807\pm0.020 \\
\Omega_m & = & 0.265 \pm 0.016. \nonumber
\end{eqnarray}
These results firmly establish optical cluster
studies as a method for deriving precise cosmological constraints.
Importantly, our results are in excellent agreement with and of
comparable precision to X-ray derived cluster abundance constraints,
clearly demonstrating the robustness of galaxy clusters as a tool
of precision cosmology.

We have discussed how and why galaxy clusters can help constrain dark
energy evolution, demonstrating that even in those data sets where the
evolution of cluster abundance with redshift is clearly detected,
constraints on the dark energy density and equation of state of a
joint WMAP and cluster abundance analysis are dominated by the low
redshift cluster normalization condition
$\sigma_8\Omega_m^\gamma=constant$.  These joint constraints are
driven by the growth of the matter fluctuations between the time of
last scattering and the low redshift universe.  Thus, while cluster
abundances provide only moderate improvements to dark energy
constraints derived from joint WMAP and supernovae analysis, we have
argued that they provide an important consistency test of general
relativity.  More specifically, our constraint on $\sigma_8$ allows us to 
rule out at the $2\sigma$ lavel any models for which the growth of structure 
between last scattering
and today differs from that of our best fit $\Lambda$CDM model by more 
than $\sim 6\%$.

At this time, the dominant systematic uncertainty in our analysis is
the uncertainty in the weak lensing mass scale due to scatter in
photometric redshift estimates of the source galaxies.  In addition,
improvements to our understanding of the scatter of the mass--richness
relation could help tighten our cosmological constraints.  Follow-up
observations can help in this regard, but the number of follow ups necessary
to have a significant impact on our results is currently very large.
Fortunately, reducing the scatter of the mass--richness relation 
\citep[see e.g.][]{reyesetal08,rozoetal08b} may 
help reduce the number of follow up observations necessary to
achieve improved constraints.

\acknowledgements We thank Hao-Yi Wu for a careful reading of the
manuscript and helpful comments.  ER was funded by the Center for
Cosmology and Astro-Particle Physics at The Ohio State University and 
by NSF grant AST 0707985. 
RHW was supported in part by the U.S. Department of Energy under
contract number DE-AC02-76SF00515 and by a Terman Fellowship at
Stanford University.  ESR would like to thank the TABASGO foundation.
TAM was funded in part by the National Science Foundation under Grant
number AST-0807304.  This work is supported in part by U.S. Department of Energy
under contract No. DE-AC02-98CH10886.  AEE acknowledges support from NSF AST-0708150.

Funding for the SDSS and SDSS-II has been provided by the Alfred P. Sloan Foundation, the Participating Institutions, the National Science Foundation, the U.S. Department of Energy, the National Aeronautics and Space Administration, the Japanese Monbukagakusho, the Max Planck Society, and the Higher Education Funding Council for England. The SDSS Web Site is http://www.sdss.org/.

The SDSS is managed by the Astrophysical Research Consortium for the Participating Institutions. The Participating Institutions are the American Museum of Natural History, Astrophysical Institute Potsdam, University of Basel, University of Cambridge, Case Western Reserve University, University of Chicago, Drexel University, Fermilab, the Institute for Advanced Study, the Japan Participation Group, Johns Hopkins University, the Joint Institute for Nuclear Astrophysics, the Kavli Institute for Particle Astrophysics and Cosmology, the Korean Scientist Group, the Chinese Academy of Sciences (LAMOST), Los Alamos National Laboratory, the Max-Planck-Institute for Astronomy (MPIA), the Max-Planck-Institute for Astrophysics (MPA), New Mexico State University, Ohio State University, University of Pittsburgh, University of Portsmouth, Princeton University, the United States Naval Observatory, and the University of Washington.

\bibliographystyle{apj}
\bibliography{mybib}

\newcommand\AAA[3]{{A\& A} {\bf #1}, #2 (#3)}
\newcommand\PhysRep[3]{{Physics Reports} {\bf #1}, #2 (#3)}
\newcommand\ApJ[3]{ {ApJ} {\bf #1}, #2 (#3) }
\newcommand\PhysRevD[3]{ {Phys. Rev. D} {\bf #1}, #2 (#3) }
\newcommand\PhysRevLet[3]{ {Physics Review Letters} {\bf #1}, #2 (#3) }
\newcommand\MNRAS[3]{{MNRAS} {\bf #1}, #2 (#3)}
\newcommand\PhysLet[3]{{Physics Letters} {\bf B#1}, #2 (#3)}
\newcommand\AJ[3]{ {AJ} {\bf #1}, #2 (#3) }
\newcommand\aph{astro-ph/}
\newcommand\AREVAA[3]{{Ann. Rev. A.\& A.} {\bf #1}, #2 (#3)}

\end{document}